\documentclass[a4paper,11pt]{article}

\usepackage[a4paper]{geometry}

\usepackage{amsmath}
\usepackage{amsfonts}
\usepackage{graphicx}
\usepackage{authblk}
\usepackage{multirow}
\usepackage{xspace} 
\usepackage[font=small,labelfont=bf]{caption}
\usepackage[colorlinks=true, linkcolor=blue, urlcolor=blue, citecolor=red, anchorcolor=green]{hyperref}
\usepackage[numbers,sort&compress]{natbib}
\usepackage{doi}

\providecommand{\keywords}[1]{\textbf{\small \textit{keywords:}} {\small #1}}

\newcommand{\Cascade}{\texttt{Cascade}\xspace}
\newcommand{\Binary}{\texttt{Binary}\xspace}
\newcommand{\BBBSS}{\texttt{BBBSS}\xspace}
\newcommand{\Winnow}{\texttt{Winnow}\xspace}
\newcommand{\BICONF}{\texttt{BICONF}\xspace}

\renewcommand{\vec}[1]{\mathbf{#1}}

\title{Demystifying the Information Reconciliation Protocol Cascade}

\author[1]{Jesus Martinez-Mateo}
\author[2]{Christoph Pacher}
\author[2]{Momtchil Peev}
\author[1]{Alex Ciurana}
\author[1]{Vicente Martin\thanks{\href{mailto:vicente@fi.upm.es}{vicente@fi.upm.es}}}
\affil[1]{\small Facultad de Inform\'{a}tica, Universidad Polit\'{e}cnica de Madrid, \authorcr
Campus de Montegancedo, 28660 Boadilla del Monte, Madrid, Spain}
\affil[2]{\small Safety \& Security Department, AIT Austrian Institute of Technology GmbH, Donau-City-Stra{\ss}e 1, 1220 Vienna, Austria}

\begin{document}

\maketitle

\begin{abstract}
\normalsize \Cascade is an information reconciliation protocol proposed in the context of secret key agreement in quantum cryptography. This protocol allows removing discrepancies in two partially correlated sequences that belong to distant parties, connected through a public noiseless channel. It is highly interactive, thus requiring a large number of channel communications between the parties to proceed and, although its efficiency is not optimal, it has become the de-facto standard for practical implementations of information reconciliation in quantum key distribution. The aim of this work is to analyze the performance of \Cascade, to discuss its strengths, weaknesses and optimization possibilities, comparing with some of the modified versions that have been proposed in the literature. When looking at all design trade-offs, a new view emerges that allows to put forward a number of guidelines and propose near optimal parameters for the practical implementation of \Cascade improving performance significantly in comparison with all previous proposals.
\end{abstract}


\section{Introduction}

Inspired by the early 1970's ideas of Stephen J. Wiesner about Quantum Money~\cite{Wiesner_83}, Quantum key distribution (QKD) emerged from the original work by Charles H. Bennett and Gilles Brassard. In 1984 they proposed the first QKD protocol \cite{Bennett_84}, commonly known as the Bennett-Brassard 1984 (BB84) protocol. However, their contribution goes beyond, and they were also pioneers in conducting the first QKD experiment as well as proposing novel procedures for the classical secret key post-processing \cite{Bennett_88, Bennett_92, Bennett_95, Brassard_94}, including information reconciliation and privacy amplification\footnote{Note that, by information reconciliation or briefly, reconciliation, we mean error correction in the context of secret key agreement.}.

Initially, in~\cite{Bennett_92} they proposed a protocol for reconciling errors based on a block parity exchange. Two correlated sequences of bit values belonging to different parties are processed in parallel. Each of the parties divide the sequence, or frame, into blocks of equal length. Then, the parity (i.e., the sum modulo 2 of all bits) of each block is computed and the respective values are exchanged through a public noiseless channel. This procedure detects all blocks with parity mismatches. For all those blocks, the parties perform a dichotomic search (a divide and conquer algorithm similar to binary search) to find and correct one of the errors that have occurred in the block. This procedure detects all blocks with an odd number of errors but corrects only exactly one error per block. Therefore, the protocol needs to work iteratively for a number of passes. In each successive pass the frame is shuffled and further parities are exchanged to detect and correct further errors. The number of remaining errors monotonically decreases with each pass, but there is no guarantee that all the errors in a frame are corrected after a number of passes. The protocol is commonly known as \BBBSS, but sometimes also referred to as \Binary.

Later in~\cite{Brassard_94} the authors realized that in \BBBSS each detected error produces side information that could be used to correct undetected errors of previous passes. Similarly, their modified protocol runs for a fixed number of passes. In each pass, the parties divide their frame into blocks of equal length. The parity is calculated and exchanged for each block, and when the parity differs the parties perform a dichotomic search to find the position of one faulty bit. For the first pass the initial block size is calculated as a function of the estimated error probability in the quantum channel or quantum bit error rate (QBER), and it is doubled for successive passes. However, since whenever an error is found after the first pass it also uncovers an odd number of additional errors masked in the preceding passes, now the algorithm steps back to correct one of them. Sometimes this correction uncovers yet another error in a different pass, starting a cascade of corrections. Therefore, this new protocol has been named \Cascade in reference to this iterative or cascading process of identifying and correcting errors in previous passes.

\Cascade is probably the most widely used and best known protocol for information reconciliation in QKD. Although it is a highly interactive protocol, requiring many communication rounds (or channel uses) between the parties (i.e., the parties have to exchange a large number of messages), it is reasonably efficient and easy to implement. Accordingly, a number of modifications and optimizations have been proposed in the literature for both, the \BBBSS and \Cascade protocols \cite{VanDijk_97a, VanDijk_97b, Yamazaki_98, Sugimoto_00, Chen_00, Furukawa_01, Yamamura_01, Chen_01, Nguyen_02, Liu_02, Liu_03, Buttler_03, Yan_08, Han_09, Ii-Yung_13}, but none of them have become as widespread. Most of these works, e.g., \cite{VanDijk_97a, VanDijk_97b, Yamazaki_98, Sugimoto_00, Chen_01, Liu_02, Liu_03, Yan_08}, concentrate on how to optimize the efficiency of reconciliation by modifying the first and subsequent block sizes. Further, some other works propose modifications to the protocol itself, for instance, combining a modified version of \Cascade with a second algorithm to improve the reconciliation efficiency \cite{Sugimoto_00}, or the number of channel communications \cite{Buttler_03}. An example of the latter modification is \Winnow \cite{Buttler_03}, and it is based on an idea initially presented in~\cite{Furukawa_01} as an improvement for \BBBSS. In both~\cite{Furukawa_01} and \cite{Buttler_03}, the authors propose to replace the dichotomic search with a linear error-correcting code (e.g., a Hamming code \cite{Hamming_50}), compute and exchange the syndrome of each block and use these to detect and correct errors reducing the number of communication rounds needed. Similarly, several works~\cite{Bennett_88, Yamamura_01}, propose to use the result of a hash function rather than the parity value to detect and correct errors in a block. Finally, other works propose combining \Cascade with an advantage distillation protocol \cite{VanDijk_97a, VanDijk_97b, Liu_02, Liu_03} and, although interesting, these are not part of the scope of this work which is mainly focused on QKD.

All the previous modifications that concentrate on \Cascade try to improve it either by optimizing the parameters in the algorithm or by modifying the protocol itself, e.g., exchanging parities of blocks obtained by another method but still keeping the idea of the cascading process. This contribution studies the possible design options in \Cascade, comparing them with the original protocol and with the most significant modifications published. The comparison is made on the grounds of a full set of parameters, so that their effects can be fairly assessed, in contrast to the limited and focused ones published up to now. Note that in the design of error correcting codes it is well known that there is not a single optimum \cite{Bonello_11}, but that a set of trade-offs have to be chosen instead. Previous modifications and improved versions of \Cascade have concentrated almost exclusively on its reconciliation efficiency, without regard to other major features. This has produced a somewhat skewed view of the real capabilities of \Cascade, hiding aspects that are important from the point of view of code design and also significant in practice. Here a number of simulations of the protocol and its most significant variants are performed to study not only the efficiency but also other characteristics that are important for its practical application, such as the number of communication rounds and the failure probability, among others. When looking at all the salient characteristics at the same time a different view emerges, showing that, for instance, an increased failure probability results from some of the supposed advantages of these modifications. This allows us to propose a set of guidelines and optimizations, which would boost its performance. Table~\ref{fig:simulation-list} summarizes all the different versions of \Cascade simulated here, their parameters, and those optimizations considered for each version. Simulated results are also analyzed considering recent studies of \Cascade \cite{Seet_13, Ii-Yung_13}, and practical implementations \cite{Pedersen_14}.

\begin{table}
\centering
\caption{Original, modified and optimized versions of \Cascade analyzed in the manuscript. A frame of length $n=10^{4}$ bits was considered for all versions of \Cascade, except for the last optimization labeled as (8) where the length of the frame used is $n=2^{14}$. The new optimizations presented in this paper are the ones labeled from (3) to (8).}
\label{fig:simulation-list}
\begin{minipage}{\linewidth}
\begin{tabular}{|l|c@{}c@{}cccccc|}
\hline
Protocol & \multicolumn{3}{c}{Block sizes (approx.)} & \Cascade & \BICONF & Block & Shuffling & Singl.\\
 & $k_{1}$ & $k_{2}$ & $k_{i}$ & passes & & reuse & & blocks \\
\hline
\begin{tabular}{@{}l@{}} orig. \\ Ref.~\cite{Brassard_94} \end{tabular} & $0.73/Q$ & $2k_{1}$ & $2k_{i-1}$ & 4 & no & no & random & no \\
\hline
\begin{tabular}{@{}l@{}} mod.~(1) \\ Ref.~\cite{Sugimoto_00} \end{tabular} & $0.92/Q$ & $3k_{1}$ & -- & 2 & yes & no & random & no \\
\hline
\begin{tabular}{@{}l@{}} opt.~(2) \\ Ref.~\cite{Yan_08} \end{tabular} & $0.8/Q$ & $5k_{1}$ & $n/2$ &10 & no & no\footnote{Although reuse of subblocks is also suggested in the optimized version of \Cascade proposed in~\cite{Yan_08}, this technique is not included in the simulation of that proposal in order to fairly compare the effect in the efficiency, communication rounds and failure probability of the suggested block sizes in~\cite{Yan_08} with the results of the original \Cascade protocol.} & random & no \\
\hline
opt.~(3) & $1/Q$ & $2k_{1}$ & $n/2$ & 16 & no & no & random & no \\
\hline
opt.~(4) & $1/Q$ & $2k_{1}$ & $n/2$ & 16 & no & yes & random & no \\
\hline
opt.~(5) & $1/Q$ & $2k_{1}$ & $n/2$ & 16 & no & yes & determ. & no \\
\hline
opt.~(6) & $1/Q$ & $2k_{1}$ & $n/2$ & 16 & no & yes & random & yes \\
\hline
opt.~(7) & $2^{\lceil \log_{2} 1/Q \rceil}$ & $4k_{1}$ & $n/2$ & 14 & no & yes & random & no \\
\hline
opt.~(8) & $2^{\lceil \alpha \rceil}$ & $2^{\lceil (\alpha + 12)/2 \rceil}$ & $n/2$\footnote{$\alpha=\log_{2}(1/Q)-\frac{1}{2}$, $k_{3}=2^{12}=4096$ and $k_{i}=n/2$ for $i>3$.} & 14 & no & yes & random & no \\
\hline
\end{tabular}
\end{minipage}
\end{table}

The rest of this paper is organized as follows. In Section~\ref{sec:pre} we introduce the information reconciliation problem, the concept of efficiency and some other definitions needed to analyze the performance of a reconciliation protocol. In Section~\ref{sec:protocol} we review the original \Cascade and some of the proposed improvements: modified versions of the protocol and optimized parameters. Then, their performance is compared in Section~\ref{sec:results}. As a result of this analysis we propose a \Cascade version that improves on the previous ones. Finally, we present our conclusions in Section~\ref{sec:conclusions}.

\section{Preliminaries}
\label{sec:pre}

Let $X$ and $Y$ be two correlated discrete random variables with binary alphabet $\mathcal{A}=\{0,1\}$ and joint probability $p_{XY}(x,y)=\Pr(X=x,Y=y)$. Note that, for convenience, we omit the random variables when there is no chance of confusion. The probability $p(x,y)$ can be also written as $p(y|x)p(x)$, such that $y$ can be seen as the output of a memoryless channel characterized by the transition probability $p(y|x)$ with input $x$. In the discrete-variable QKD case, errors or discrepancies between variables $x$ and $y$ belonging to two distant parties, Alice and Bob, respectively, are assumed to be the consequence of a transmission over a binary symmetric channel with crossover probability $\epsilon$, BSC($\epsilon$). The channel parameter $\epsilon$ is usually referred to as quantum bit error rate $Q$ or QBER.

Let the sequences $\vec{x} \in \mathcal{A}^{n}$ and $\vec{y} \in \mathcal{A}^{n}$ be the outcomes of $n$ independent and identically distributed (i.i.d.) instances of $X$ and $Y$, respectively. Note that, hereinafter, we refer to these sequences as frames. The problem of reconciliation is equivalent to a particular case of source coding with side information, also known as Slepian-Wolf coding \cite{Slepian_73}. Given a source $X$ and a decoder with access to side information $Y$, no encoding of $X$ shorter than $H(X|Y)$ allows for a reliable decoding in the receiver \cite{Slepian_73}. Thus, the minimum information is given by the conditional entropy $H(X|Y)$. Let $m$ be the length of the message exchanged for reconciling the discrepancies between $\vec{x}$ and $\vec{y}$. Then the efficiency of an information reconciliation procedure can be defined as:

\begin{equation}
\label{eq:efficiency}
f_{EC} = \frac{m}{nH(X|Y)}.
\end{equation}

Since $nH(X|Y)$ is the minimum length of the message transmitted to reconcile the frames $\vec{x}$ and $\vec{y}$, we have that $f_{EC} \ge 1$, and $f_{EC}=1$ stands for perfect reconciliation.

In the case of a BSC($\epsilon$) the reconciliation efficiency can be written as:

\begin{equation}
\label{eq:efficiency-bsc}
f_{EC} = \frac{1-R}{h(\epsilon)}
\end{equation}

\noindent where the binary Shannon entropy $h(\epsilon) = -\epsilon \log_{2} \epsilon - (1-\epsilon) \log_{2} (1-\epsilon)$, and $R$ is the ratio of information transmitted, $R=1-m/n$. The difference $1-R$ is the ratio of redundant information disclosed for reconciling errors.

Note that a different interpretation for the reconciliation efficiency is often used in the literature. While we have defined it as a measure of the percentage of additional information disclosed over the Shannon limit, in other works the efficiency is defined as the ratio of the capacity achieved for a given communication channel. This other value for the reconciliation efficiency is then given by:

\begin{equation}
\label{eq:efficiency-beta}
\beta = \frac{R}{1-h(\epsilon)},
\end{equation}

\noindent such that

\begin{equation}
1 - f_{EC} h(\epsilon) = \beta (1 - h(\epsilon)).
\end{equation}

Throughout this contribution we only use the first definition, but ultimately we also provide some values for the second one in order to compare our results with results presented elsewhere.

In addition, any error-correcting method has to be analyzed taking also into account its robustness. We use two measures for robustness: (i) the failure probability or frame error rate, here denoted by $\varepsilon_{EC}$, is the probability that after reconciliation the frames belonging to both parties differ by at least one bit, (ii) the residual error or bit error rate is the ratio between the number of different bits in both frames after the reconciliation process and the frame length $n$. Note also that, hereafter, we use the terms frame error rate and bit error rate rather than failure probability and residual error, respectively.

\section{\Cascade protocol and modifications}
\label{sec:protocol}

For a practical consideration of this work and a proper interpretation of the results in the next section, we provide first a detailed description of \Cascade, thus including our interpretation of some points not described in the original and modified versions of this protocol.

Then, we discuss the possible modifications of \Cascade, but only consider those methods that preserve the iterative parity exchange procedure that gives name to the protocol. The modifications considered are classified either as protocol modifications, when the rules applied to the iterative process differs from the original ones, or as protocol optimizations, when different parameters are proposed or when a particular interpretation of the protocol differs. For each case we select the main representatives in the literature and analyze its behavior in Section~\ref{sec:results} as a basis to propose a set of rules that allow to optimize \Cascade under all situations. Care has been taken to study all the relevant magnitudes and not to concentrate on just one single aspect, as it has been the case in many of the previous studies.

\subsection{The original protocol}
\label{sec:original-protocol}

As described above, \Cascade works in successive passes. Let $k_{i}$ be the block size used in the $i$-th pass of the algorithm. In the first pass, the parties divide their frames into blocks of equal length. The block size $k_{1}$ of the first division is agreed upon by both parties and calculated as a function of $Q$, the QBER. As suggested in~\cite{Brassard_94, Crepeau_95} $k_{1} \approx 0.73/Q$ is used. In particular, for the results labeled as the original \Cascade below, we used the smallest integer greater than or equal to this approximation, i.e., $k_{1} = \lceil 0.73/Q \rceil$. Then, the parties compute a parity per block, exchange this through a public noiseless channel, and perform a dichotomic search\footnote{Both parties perform the following steps: (i) divide the block into two halves, (ii) calculate the parity of the first half, and (iii) exchange that parity. If Alice and Bob obtain different parities, a bit error has to be in the first half and they continue their bisection and parity exchange there. If they obtain the same parity for the first half, a bit error must be in the second half and they continue their bisection and parity exchange there. In this way they continue until they have located the exact position of a bit error in at most $\lceil \log_{2} k_{1} \rceil$ steps.} if their parity values differ. However, note that in a practical implementation of \Cascade, blocks and parities are processed in parallel. Therefore, instead of exchanging messages with single parities typically a set of parities (i.e., a syndrome) are processed and communicated. In what follows, all the non-dependent information is collected in one message until the protocol can no longer proceed and the message is transmitted. Our results show then the minimal number of messages needed. Note that dichotomic searches (i.e., subblock parities) are also processed in parallel.

In each following pass the block size is doubled, $k_{i}=2k_{i-1}$, and the process of exchanging parities and correcting errors is repeated. From the second pass onward, each detected error can be used to correct further errors in other already completed passes. For instance, suppose that an error is detected during the second pass. This means that during the first pass this bit error was inside a block $\mathcal{B}_{1}$ with an even number of errors, and has thus remained undetected. Consequently, there must be a second error in $\mathcal{B}_{1}$ that can now be corrected. The cascade process begins always from the first pass onward to correct as many errors as possible disclosing the minimum number of parities required by a dichotomic search. Note that in the original description of \Cascade \cite{Brassard_94}, errors (i.e., discrepancies between the frames to reconcile) are assumed to be i.i.d., such that no random shuffling is proposed prior to the first pass, but in order to detect new errors the frame is randomly shuffled between the following passes. Note also that, after the end of each pass an even number of errors (possibly zero) remains in the frame; thus the parity of the last block is determined by the parities of all the previous blocks and need not be exchanged from the second pass onward. This is similar to the dichotomic search in which only the parity of the first half needs to be exchanged. Also, from an information leakage point of view, in the second and all following passes the last block's parity is redundant and need not be taken into account in the calculation of the protocol's leakage. Finally, the protocol concludes when four passes have been completed. As suggested in~\cite{Brassard_94}, these four passes have proved to be empirically enough to remove all discrepancies in a frame of length $10^{4}$ bits.

\subsection{Modified protocols}
\label{sec:modifications}

Most of the modified versions of \Cascade, e.g., \Winnow \cite{Buttler_03}, involve the substitution of the parity exchange by the use of a one-step (forward) error correcting method. However, this approach is not compatible with the iterative parity exchange process described above. The only one exception in this respect is discussed in~\cite{Sugimoto_00}. In this modification, after the first two passes of the original \Cascade, the iterative process continues with a different algorithm referred to as \BICONF. This algorithm is a slightly different version of the one already proposed in~\cite{Brassard_94} with an identical name.

In~\cite{Sugimoto_00} the block sizes used for the first two passes of \Cascade, $k_{1}$ and $k_{2}$, respectively, are given by:

\begin{equation}
k_{1} = \left \lfloor \frac{4 \ln 2}{3Q} \right \rfloor \approx 0.92/Q, \qquad
k_{2} = \left \lfloor \frac{4 \ln 2}{Q} \right \rfloor \approx 3k_{1}.
\end{equation}

These values have been derived from the observation that, after the first and second passes, approximately $p_{1}=50\%$ and $p_{2}=100\%$ of the bit errors have been corrected, respectively. Assuming $p_{1}=1/2$ and $p_{2}=1$, the proposed values should minimize the number of exchanged parities, thus optimizing the reconciliation efficiency. Note that, compared to the values in the original \Cascade (see Section~\ref{sec:original-protocol}), this suggests that the number of errors corrected during the first two passes is now lower than in the original protocol (it can be shown that the probability to correct errors in the first pass strictly decreases with increasing block size $k_{1}$).

After these two steps, the iterative \BICONF algorithm is executed. It works as follows: In each iteration, first, the parties agree on a random subset of bits from their frames. Then, they compute and exchange the parity value of this subset, and perform two dichotomic searches if their parities differ, one for the chosen subset and the other for the complementary subset (i.e., the subset of bits that were not selected). The algorithm chooses new random subsets of bits in each iteration, and stops when it has either performed $s$ iterations~\cite{Brassard_94}, or $s$ successive iterations without finding new errors~\cite{Sugimoto_00}. We consider here the latter choice to be $s=10$. Note that the process of choosing the random subset of bits is not specified. We decided to choose it by performing independent Bernoulli processes with success probability one half for each bit of the frame. This divides the frame into two subsets (a chosen subset and its complement w.r.t.\ the frame) of similar size.

In Section~\ref{sec:results} it is shown that this modified version improves the efficiency of \Cascade. However, the extensive simulations performed show that the frame error rate is considerably higher in this protocol than in the original \Cascade. Therefore, the efficiency improvement comes at the cost of a higher frame error rate---a fact that is typical for one-way reconciliation with block codes. The results in Section~\ref{sec:results} also highlight that, as already shown in~\cite{Yan_08}, one pass of \Cascade with a block size equal to one half of the frame length (i.e., $k_{i}=\lceil n/2 \rceil$) works effectively as one iteration of \BICONF, but with the advantage of possibly correcting further errors in previous passes.

\subsection{Other runtime optimizations and \Cascade parameters}
\label{sec:opt-parameters}

In the following we chronologically describe four possible optimizations of \Cascade that have been previously proposed in the literature, but are implemented and analyzed together for the first time here: (i) improving the shuffling between passes, (ii) removing \textit{singleton} blocks after each pass, (iii) optimizing block sizes, and (iv) reusing subblocks resulting from dividing the frame in the dichotomic search. The efficiency for all these optimized versions of \Cascade is discussed later in Section~\ref{sec:results} on the basis of extensive simulations that we have carried out.

Originally, two novel ideas for optimizing \Cascade have been put forward in~\cite{Chen_00, Chen_01}. In an unpublished draft, the author suggests that the protocol can be optimized by improving the random shuffling between passes and discarding \textit{singleton} blocks in successive passes. By singleton the author refers to a subblock of size one for which the value is known\footnote{Either because it has been exchanged or can be deduced from other, previously known, subblock parities.}. Note that then the length of the frame to reconcile decreases with each pass of the algorithm, and the block sizes suggested for other proposals are probably not optimal for this. However, the author of~\cite{Chen_01} fails to propose any method for improving the shuffling. The first practical description of a modified shuffling is proposed in~\cite{Nguyen_02}. Instead of using a random shuffling between passes, the author of this publication proposes two methods to deterministically distribute the bits of a block in a pass into different blocks in the following pass, in an attempt to uniformly distribute the errors in successive passes.

A different avenue is to leave the \Cascade protocol unchanged, but modify its parameters instead. Some optimized block sizes are also suggested in~\cite{Nguyen_02} and a comprehensive search for the optimal parameter set was later done in~\cite{Yan_08} for frames of length $10^{4}$ bits. Different values for the first block size and its subsequent size increments were analyzed. The optimal efficiency of \Cascade was empirically determined to occur for $k_{1}=0.8/Q$, $k_{2}=5k_{1}$ and $k_{i}=n/2$ for $3 \le i \le 10$. Unfortunately, as in the case before (see previous Subsection), the size of the simulation was neither large enough to determine the frame error rate nor was this aspect taken explicitly into account to produce a unskewed comparison with the original \Cascade.

Finally, another novel improvement of \Cascade is also proposed in~\cite{Yan_08}. It was emphasized that, according to the original description of \Cascade, the protocol only considers blocks resulting from dividing the frame (i.e., the blocks of size $k_{i}$ for the $i$-th pass) at the beginning of each pass. However, in a proper interpretation, also those blocks resulting from the dichotomic search can be reused. The protocol can take advantage of the smaller blocks for correcting errors during the cascade process disclosing fewer parities. As shown below, a comprehensive record of all processed blocks leads to a further improvement in efficiency.

\section{Simulation results}
\label{sec:results}

Simulation results were computed for the original \Cascade protocol \cite{Brassard_94} and the modified and optimized versions proposed in~\cite{Sugimoto_00, Chen_01, Nguyen_02, Yan_08}. Initially, the original \Cascade is compared to the modified protocol described in~\cite{Sugimoto_00}, that uses two passes of \Cascade and subsequent iterations of \BICONF (see subsection~\ref{sec:modifications}). Then, a modification of~\cite{Sugimoto_00} is proposed by replacing \BICONF for a number of passes of \Cascade with block size half of the frame length, as already hinted in~\cite{Yan_08}, but using the first block size suggested as optimal in our previous simulations. Results are later compared to a version using the block sizes suggested in~\cite{Yan_08}. Finally, those novel optimizations described in~\cite{Chen_01, Nguyen_02, Yan_08} are also considered, and a fully optimized version is presented for the first time.

Simulations were performed that cover the full error range of interest in BB84 using as a base frame length $n=10^{4}$ bits. We consider this value to be a good choice, given that hardware implementations are feasible for this size but become problematic for bigger sizes (e.g., due to physical memory limitations of FPGAs). This frame length was also used in~\cite{Brassard_94, Sugimoto_00, Chen_01, Nguyen_02, Yan_08}, which allows for a fair comparison between these proposals as well as our results. Other frame lengths (from $n=10^{3}$ to $n=10^{6}$) have been used whenever necessary. For all simulations, correlated pairs of random bit frames were generated using a congruential pseudo random number generator with a common (previously shared) seed. Given the channel parameter $Q$ (i.e., the quantum bit error rate, QBER), errors were generated in one of the frames simulating independent Bernoulli processes with success probability $Q$. Reconciliation efficiency, communication rounds, frame error rate and bit error rate have been exhaustively computed for each version considered in order to ensure a fair comparison. For instance, to analyze the latter two quantities we have simulated more than $10^{6}$ frames for all values reported here.

\subsection{Original and modified versions of \Cascade}

\begin{figure}[ht]
\centering
\includegraphics[width=0.8\linewidth]{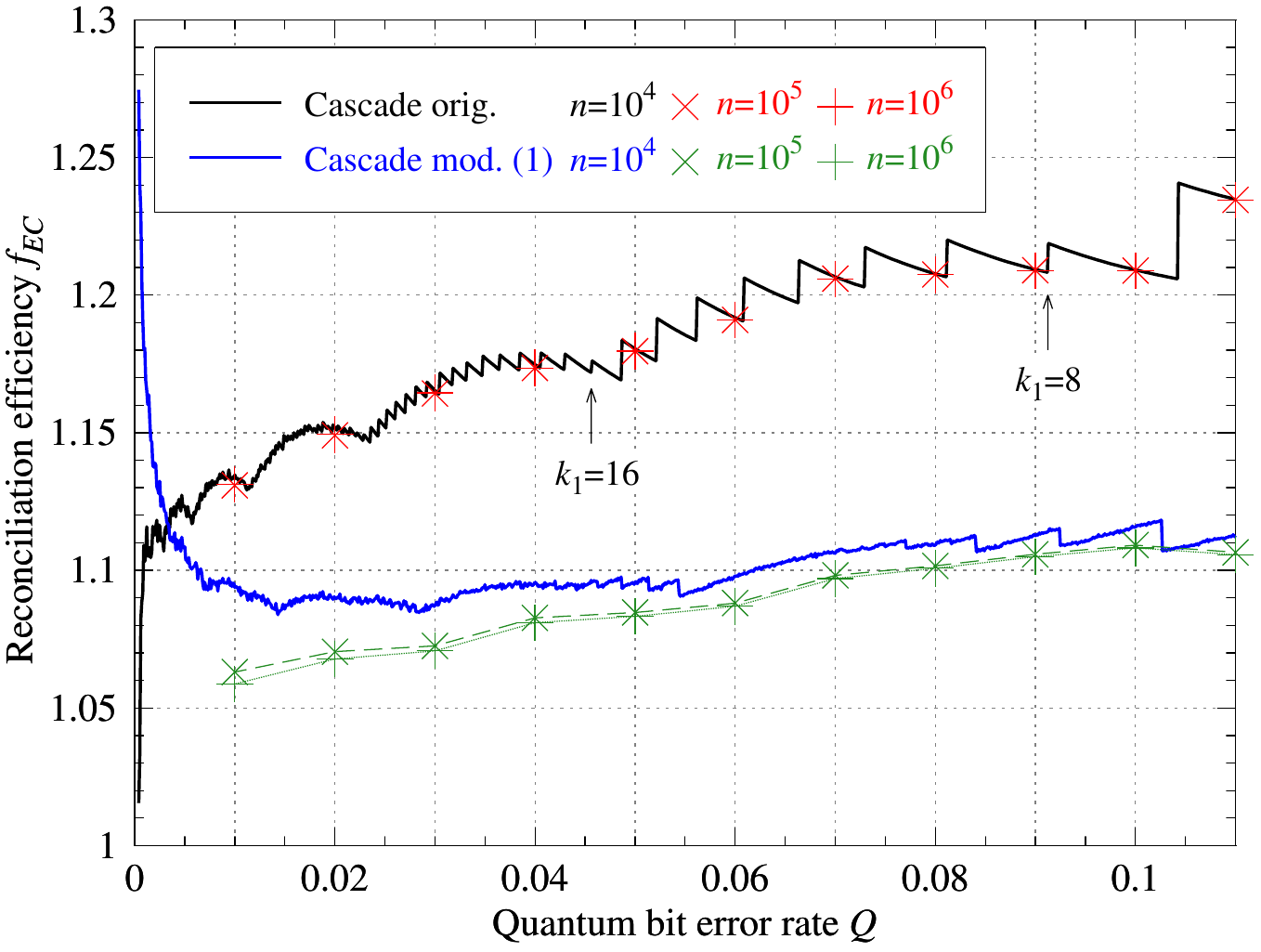}
\caption{Comparison of the average reconciliation efficiency as a function of the quantum bit error rate $Q$ for the original \Cascade protocol~\cite{Brassard_94} and the version of~\cite{Sugimoto_00} (i.e., modified version (1) throughout this paper). The length of the frames used is denoted by $n$.}
\label{fig:efficiency}
\end{figure}

\begin{figure}[ht]
\centering
\includegraphics[width=0.8\linewidth]{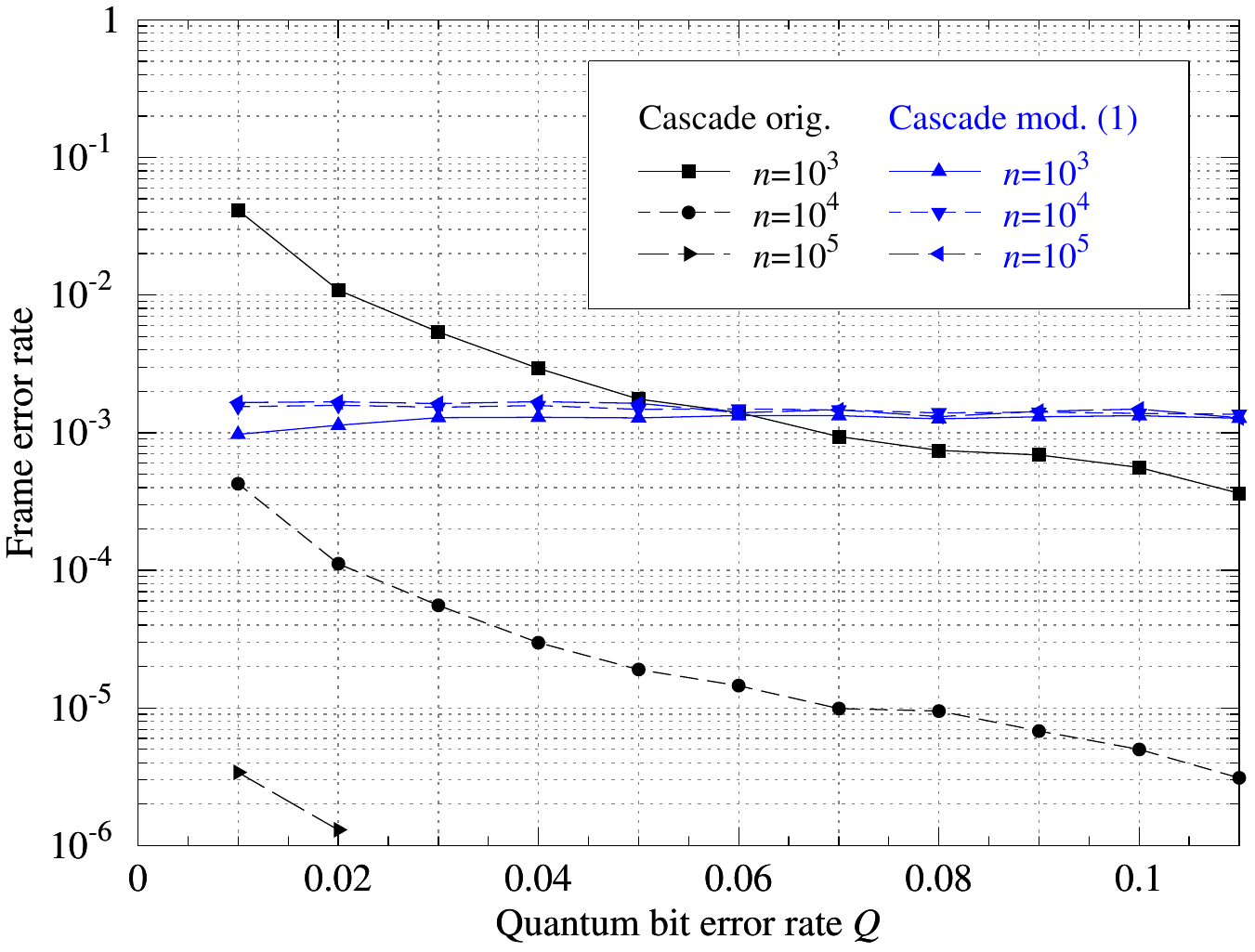}
\caption{Frame error rate (failure probability) as a function of the quantum bit error rate for the original \Cascade protocol~\cite{Brassard_94} and the version (mod.~(1)) of~\cite{Sugimoto_00}.}
\label{fig:frame-error-rate}
\end{figure}

Fig.~\ref{fig:efficiency} shows the average reconciliation efficiency as a function of QBER for the original \Cascade \cite{Brassard_94} and the modified version proposed in~\cite{Sugimoto_00}. Efficiency is calculated as defined in Eq.~(\ref{eq:efficiency-bsc}). This figure shows that the efficiency of the modified version of \Cascade improves for this frame length when the error rate is greater than approximately $0.5\%$. However, the efficiency of both protocols in the region of QBER below $\approx 1\%$ is not directly comparable because they have completely different frame error rates, as shown in Fig.~\ref{fig:frame-error-rate}. Results for longer frames have also been computed, but for a much smaller number of error rates. For these error rates, \Cascade's efficiency does not improve for longer frames while it does, although marginally, for the modified version. Therefore, a first strength or weakness (arguably) of \Cascade to be highlighted is that short frames can be corrected as efficiently as longer ones. On the other hand, modified versions of \Cascade may slightly improve the efficiency by increasing the length of the frame to reconcile. The efficiency curves for both protocols exhibit a sawtooth behavior due to the discreteness of the block sizes $k_{i}$, $1 \le i \le 4$ (jumps occur at those values of $Q$ where $k_{1}$ changes its integer value, and subsequently $k_{2}$, $k_{3}$ and $k_{4}$). Some of these $k_{1}$ values are marked in the figure. Note that, for instance, the point marked as $k_{1}=8$ is the value for which the protocol decrements the first block size from 9 to 8. Thus, a block size of 9 bits is used for the first pass in the region immediately to the left of that point, and blocks of size 8 to the right. This reduction in the block sizes directly affects the reconciliation efficiency since the number of blocks per frame increases, hence the number of disclosed parities. As shown below, the large jumps arise from a poor choice of the initial block size $k_{1}$.

\begin{figure}[ht]
\centering
\includegraphics[width=0.8\linewidth]{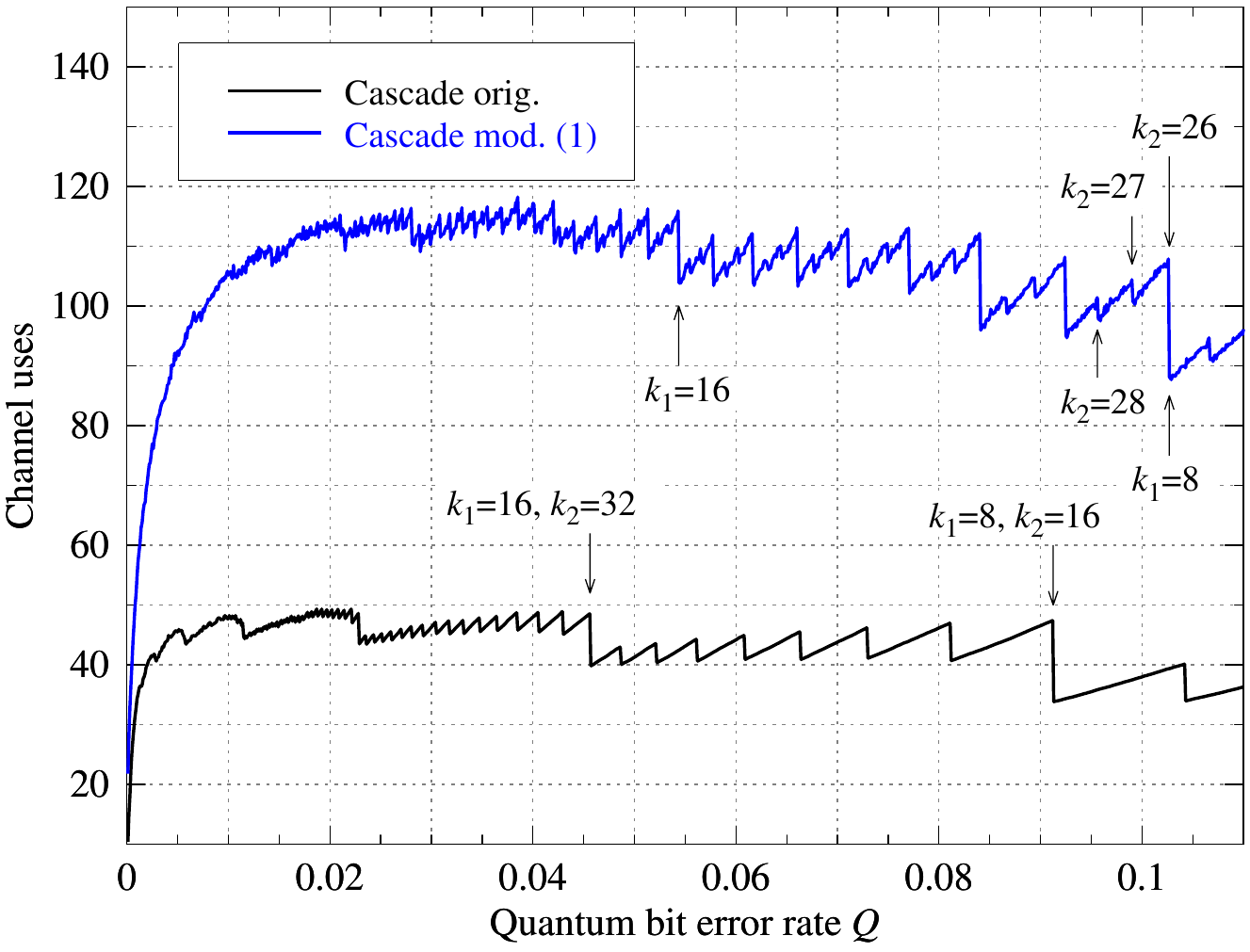}
\caption{Comparison of the number of channel uses as a function of the quantum bit error rate $Q$ for the original \Cascade protocol~\cite{Brassard_94} and the version (mod.~(1)) of~\cite{Sugimoto_00}.}
\label{fig:channel-uses-Q}
\end{figure}

The same simulations that we present in Fig.~\ref{fig:efficiency} are also used in Fig.~\ref{fig:channel-uses-Q} to compare the number of channel uses required by \Cascade and its modified version. By channel uses we mean the number of communication rounds or pair of messages exchanged through the noiseless channel to disclose parity values\footnote{Note that we do not consider other uses of the communication channel, such as the messages exchanged to synchronize the frame shuffling. Note also that we consider just one use of the channel although two messages are exchanged between the parties at once (i.e., simultaneously), each one traveling in opposite directions \cite{Momtchil_14}.}. Fig.~\ref{fig:channel-uses-Q} shows the number of channels uses as a function of QBER for frames of length $10^{4}$ bits. As shown, in this case the price to pay for improving the reconciliation efficiency is an increase (a significant one, more than double) in the number of channel uses. However, later we show that this is not entirely true, since the frame error rate has also to be taken into account (see Fig.~\ref{fig:frame-error-rate} below). As in the previous figure, the curves for both protocols also exhibit a sawtooth behavior due to the discreteness of the block sizes. Some of the respective $k_{1}$ and $k_{2}$ values are also marked. The effect of $k_{2}$ is also clearly to be noticed for the modified version of \Cascade as a smaller amplitude sawtooth behavior seen for the same value of $k_{1}$. Note however that the effect with respect to the communication rounds is the opposite to the one observed in the efficiency analysis: the number of communications decreases when the block sizes also decreases. As shown in the curve for the modified version of \Cascade, this effect due to changes in $k_{1}$ is more pronounced compared to changes in the other block sizes.

\begin{figure}[ht]
\centering
\includegraphics[width=0.8\linewidth]{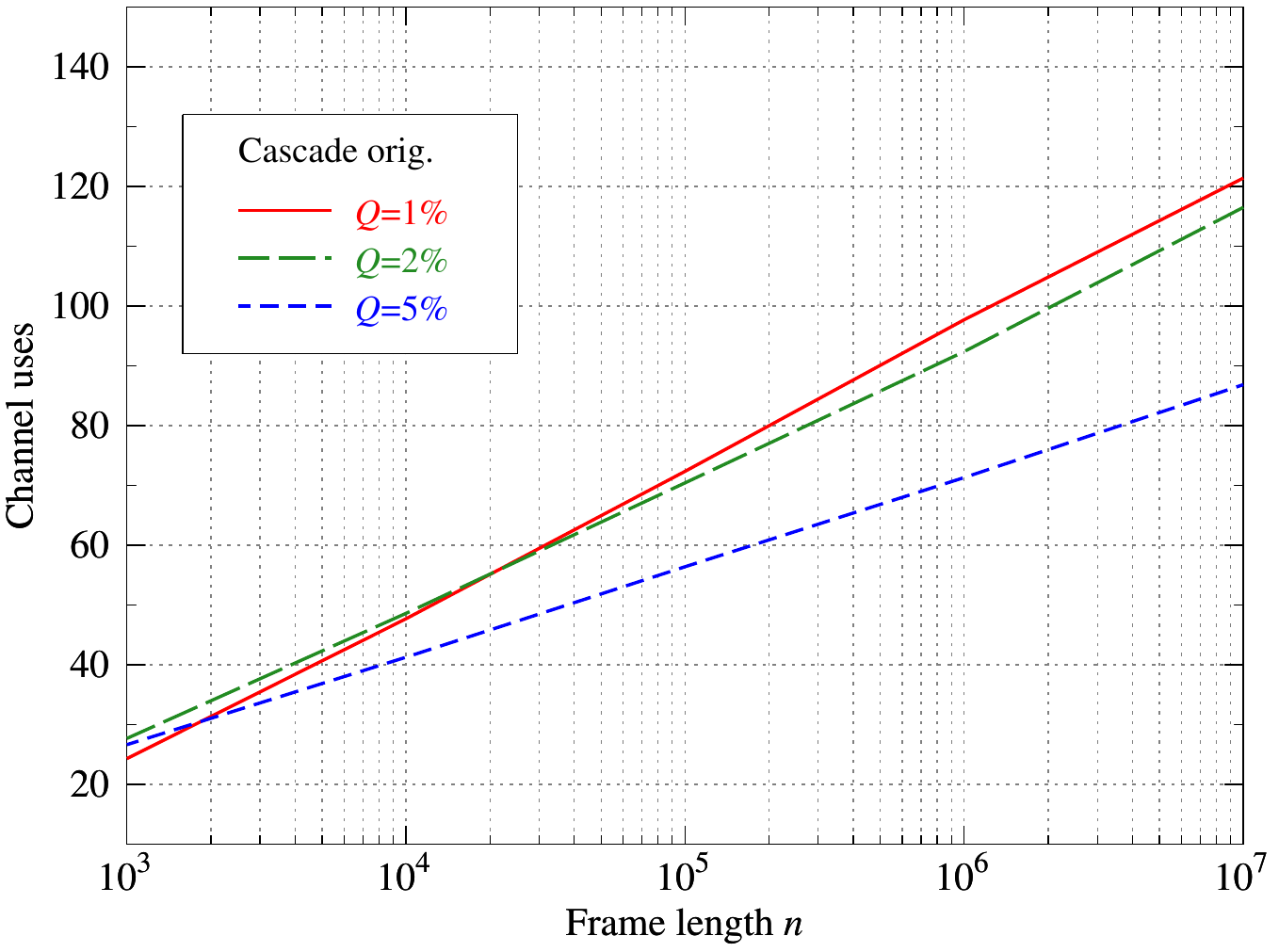}
\caption{Channel uses (communication rounds) as a function of the frame length $n$ for the original \Cascade protocol~\cite{Brassard_94}.}
\label{fig:channel-uses-N}
\end{figure}

In Fig.~\ref{fig:channel-uses-N} channel uses are shown as a function of the length of the frame to reconcile. Only the original \Cascade is considered here. The number of channel uses are computed by increasing the frame length for a constant QBER value $Q$. Three different values, $Q=1\%$, $2\%$ and $5\%$, are considered in the figure. As it was already shown in~\cite{Pedersen_14}, the number of channel uses is an increasing function of the frame length. However, here we also show results for shorter frames than those considered in~\cite{Pedersen_14}. Note that the number of channel uses depends on the frame length, block sizes and QBER. On the one hand, for higher QBER values the block sizes decrease, and accordingly the depth of the binary search tree also decreases, thus the protocol would require less channel uses. On the other hand, the number of channel uses should increase with both the frame length and the QBER due to the effectiveness of error backtracking. Indeed, as shown below, the frame and bit error rates decrease in the original \Cascade for larger frames and higher QBER values, which happens at the cost of an increased number of communication rounds. The net behavior, resulting from these partly conflicting tendencies is illustrated in Fig.~\ref{fig:channel-uses-N}.

Fig.~\ref{fig:frame-error-rate} shows the frame error rate as a function of QBER. Again, results are shown for the original \Cascade \cite{Brassard_94} and the modified version proposed in~\cite{Sugimoto_00}. Note that the frame error rate does not take into account the number of erroneous bits at the end of the protocol. Contrary to what happens with other reconciliation methods, when \Cascade ends there is no validation method to determine whether the protocol could have failed\footnote{For instance, when working with linear codes, the syndrome of the decoded word confirms whether it corresponds to a codeword. In this case, we assume that the decoding was successful although there is still a non-zero probability of having undetected errors. Otherwise, it is known that the decoding process failed. However, note that in QKD post-processing information reconciliation is always followed by a validation phase, which guarantees that the maximal frame error rate is below a certain value, whereby the latter can be chosen at will and is part of the overall security figure of merit of the protocol.}. As shown, the frame error rate is significantly higher for the modified version of \Cascade. Therefore, although the efficiency improves, the fraction of successfully reconciled frames worsens. Different frame lengths have been considered and compared, and it is evident that while the frame error rate decreases with the frame length in \Cascade, this is not the case for the modified version, for which for lengths of $10^{5}$ bits the frame error rate remains remarkably constant at $10^{-3}$.

Note that with respect to this parameter we have found significant discrepancies with other results published in the literature. In~\cite{Sugimoto_00} the authors reported that ``the modified protocol never failed''. However, they only simulated one hundred frames, which is clearly not enough to empirically verify frame error rates of the order of $10^{-3}$, let alone assume that this is a fair comparison with the original \Cascade. A similar behavior, i.e., with zero frame error rate, was also reported in~\cite{Yan_08}, even though the frame error rate of this protocol was previously known not to be negligible. For instance, in~\cite{Chen_01} the author suggests that a frame error rate of approximately $10^{-6}$ is achieved for frames of length $10^{4}$ bits\footnote{We have not confirmed this result since it is reported for a QBER of $15\%$, a value completely out of the scope of the present work}. A significant problem in the interpretation of these results arises from the fact that the small number of simulated frames also affects the efficiency and produces, spuriously, better values than those shown here. Having good statistics is extremely important in order to have a precise efficiency value: the error probability in the last passes of \Cascade (e.g., passes 3 and 4) is known to be low and if not enough samples are used, an effective zero frame error rate might be found. As an example of this effect, we have performed two simulations of a hundred frames for $Q=2\%$ and $Q=5\%$ with zero frame error rates, the efficiencies obtained are $f_{EC}=1.08171$ and $1.09264$, respectively; while the ``real'' efficiencies, based on good statistics, are slightly worse being $f_{EC}=1.09013$ and $1.09541$ for $2\%$ and $5\%$ of QBER, respectively.

\begin{figure}[ht]
\centering
\includegraphics[width=0.8\linewidth]{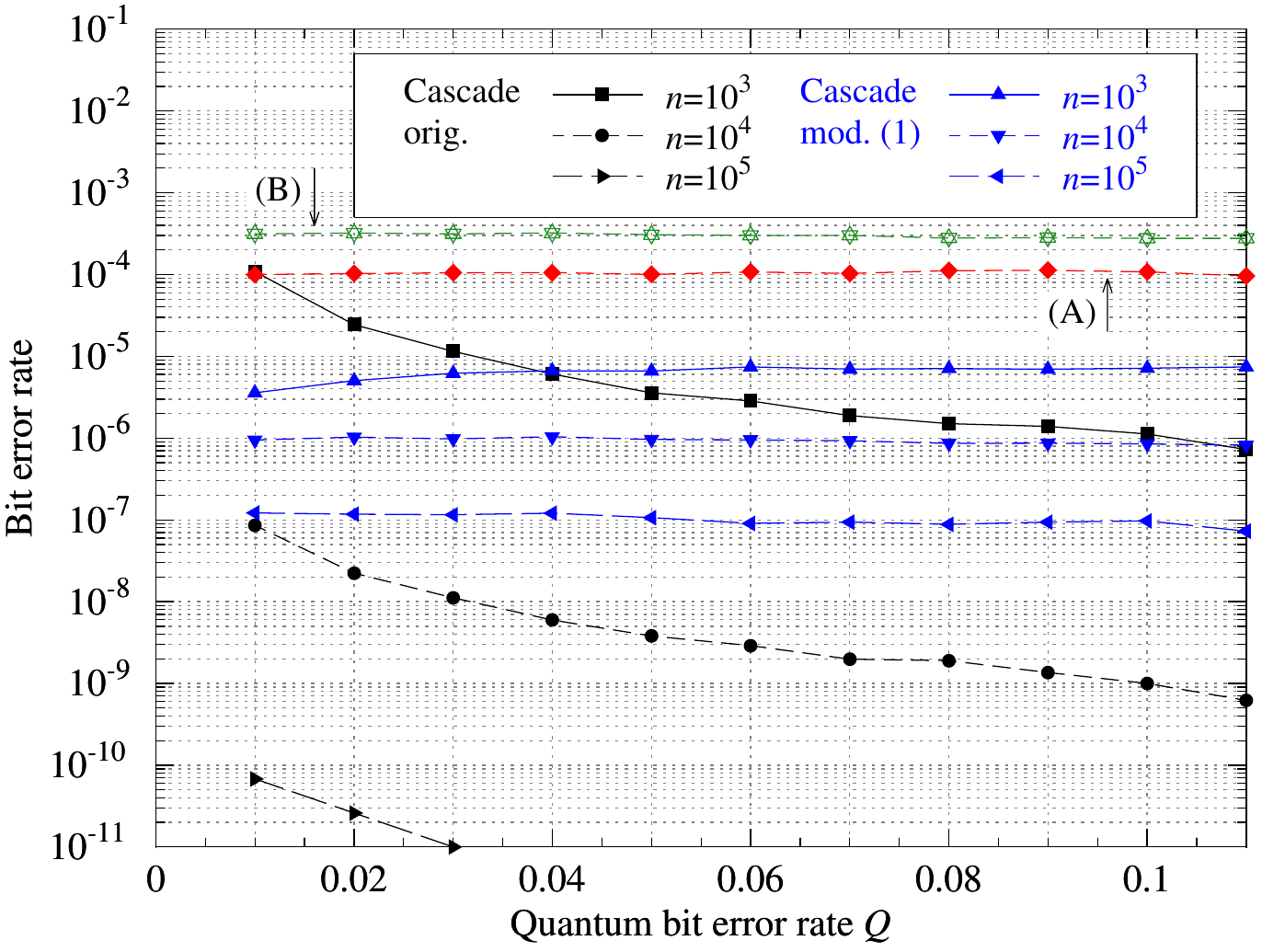}
\caption{Bit error rate (residual error) as a function of the quantum bit error rate for the original \Cascade protocol~\cite{Brassard_94} and the version (mod.~(1)) of~\cite{Sugimoto_00}. The curves labeled (A) and (B) correspond to the bit error rate after the first two passes of the original \Cascade (A) and two passes in the modified one without \BICONF (B).}
\label{fig:bit-error-rate}
\end{figure}

Finally, Fig.~\ref{fig:bit-error-rate} shows the bit error rate (see Section~\ref{sec:pre} for the definition) as a function of QBER. Unlike in Fig.~\ref{fig:frame-error-rate}, this ratio reflects the number of errors remaining in the frame at the end of the protocol. As in the previous figures, simulation results computed for the original and modified \Cascade, and different frame lengths are presented. In the figure, two additional curves are included, labeled as (A) and (B), corresponding respectively to the bit error rate after the first two passes of the original \Cascade and two passes in the modified one without \BICONF. It is seen that \Cascade works better due to its capability of tracing back extra errors. Later we use this bit error rate as an estimate of the suitability of the third and subsequent block sizes.

\subsection{Simulating \Cascade as a rateless protocol}

In this section we study the ability of the protocol to adapt to variations in the communication channel, i.e., the rateless behavior of \Cascade is analyzed. To this end simulations have been carried out using two different input parameters instead of only one. We varied (i) the error rate value $p$ used to initialize the protocol, i.e., the first block size $k_{1}$ is now derived from $p$ and not from $Q$; and (ii) $Q$ the actual quantum bit error rate, i.e., the error rate value used to generate discrepancies in the correlated frames. Note that $p$ may stand for a (poor) estimate of $Q$. Therefore, the following simulations show how the protocol behaves under time-varying channel conditions. In addition, as discussed below, these simulations give more insight about some parameters used in the protocol (e.g., block sizes) and suggests possible optimizations. We remind that the first block size $k_{1}$, and consequently the subsequent block sizes, are chosen depending on the QBER estimate.

\begin{figure}[ht]
\centering
\includegraphics[width=0.8\linewidth]{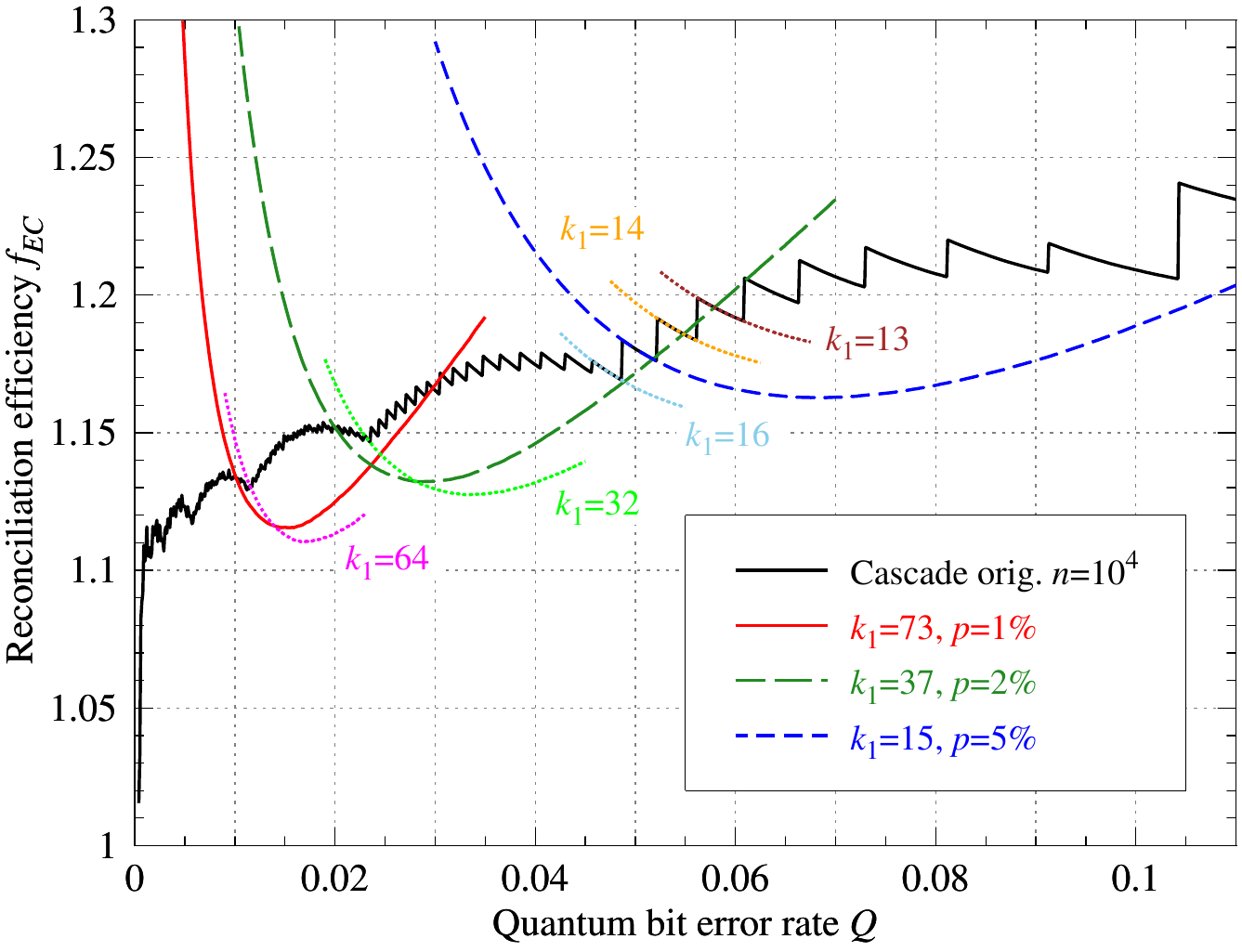}
\caption{Average reconciliation efficiency, $f_{EC}$, as function of the quantum bit error rate when $k_1$ is fixed over a larger interval of $Q$ than originally proposed~\cite{Brassard_94}. For comparison also the efficiency of the original \Cascade protocol~\cite{Brassard_94} is shown.}
\label{fig:efficiency-rateless}
\end{figure}

Fig.~\ref{fig:efficiency-rateless} shows the average reconciliation efficiency as a function of QBER. Three different cases have been considered using a constant estimate of $p=1\%$ (red), $p=2\%$ (green), and $p=5\%$ (blue), respectively for the initialization of \Cascade. Following the description of the original \Cascade protocol~\cite{Brassard_94} (see Section~\ref{sec:original-protocol}) results have been computed for frames of bit length $n=10^{4}$. In this case we get $k_{1}=73$, $k_{1}=37$, and $k_{1}=15$, respectively. The efficiency of an unmodified \Cascade is also depicted in the figure, and as expected, it coincides with the new simulations whenever $Q=p$. Interestingly, it is shown that the efficiency improves in a range of QBER values greater than the error rate considered for the initialization, i.e., for $Q>p$. This improvement coincides with the decreasing segments that produce a sawtooth shaped efficiency curve, as shown in the figure for $k_{1}=13$, $14$, $15$ and $16$. Curves for $k_{1}=32$ and $k_{1}=64$ are also depicted to show that these values coincide with local minima in the global efficiency curve of the original \Cascade. Apparently, these results suggest that larger block sizes must be considered for the first block size $k_{1}$.

\begin{figure}[ht]
\centering
\includegraphics[width=0.8\linewidth]{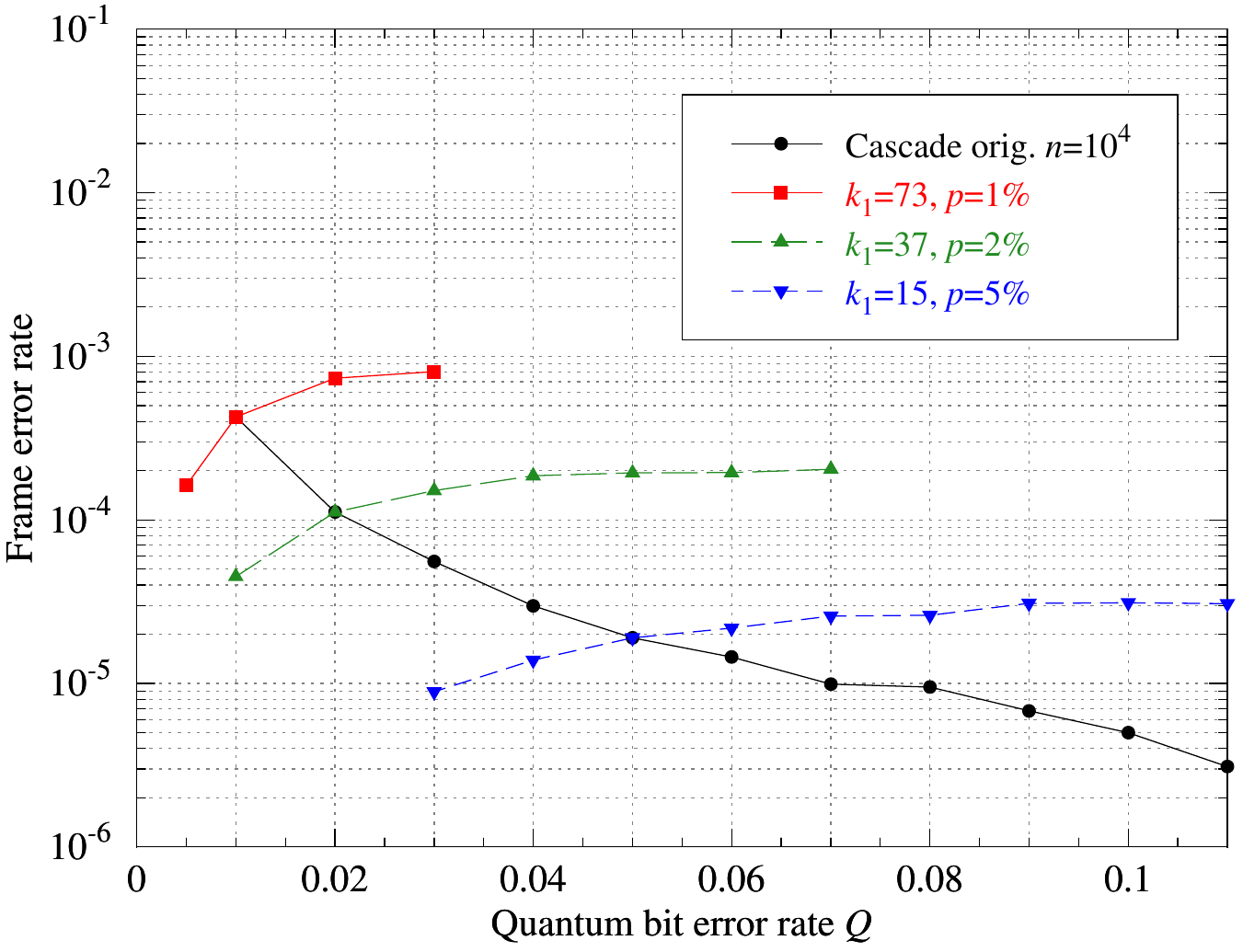}
\caption{Frame error rate (failure probability) as a function of the quantum bit error rate when $k_1$ is fixed over a larger interval of $Q$ than originally proposed~\cite{Brassard_94}. For comparison, also the frame error rate of this original \Cascade protocol is shown.}
\label{fig:frame-error-rate-rateless}
\end{figure}

As expected, a price to pay for a better reconciliation efficiency is a sharp increase in the number of exchanged messages (not shown), due to more errors being detected and corrected during the later algorithm passes. However, in the subsequent Fig.~\ref{fig:frame-error-rate-rateless} it is shown that ---surprisingly and contrary to what one might expect from the above results--- the frame error rate is not significantly affected: while the efficiency reaches its optimum for $Q>p$, the frame error rate has still the same order of magnitude. In consequence, these results clearly show that the efficiency of the original \Cascade protocol can be improved just by updating the initial block size, $k_{1}$, without modifying the rest of the protocol, and only penalizing its practical use in high latency networks due to an increased interactivity.

Thus, we empirically show that the efficiency of the original \Cascade is optimal for the three cases $p=1\%$, $2\%$ and $5\%$ when $Q \approx 1.46\%$, $2.85\%$ and $6.87\%$, respectively. Taking into account that the frame error rate does not significantly increase, and disregarding the channel uses, it follows that (for frames of bit length $n=10^{4}$) the block size $k_{1}=73$ is optimal when $Q=1.46\%$ (i.e., $k_{1}=1.0658/Q$). For the three cases considered we get that the criterion $k_{1} \approx 1/Q$ is presumably optimal. In other words, the simulated results suggest using as the first block size the value that divides the frame into blocks with one error on average. Note that this is a criterion that was very recently also suggested in~\cite{Ii-Yung_13}, and according to the evidence, it tries to maximize the number of errors corrected disclosing the minimum number of parities during the first pass.

\subsection{Some protocol optimization guidelines}

From the above results we can infer some guidelines that may be useful in finding the optimal block sizes for \Cascade: (i) first, from Figs.~\ref{fig:efficiency-rateless} and~\ref{fig:frame-error-rate-rateless} it seems that the size for the initial block should be slightly larger than the one proposed in the original protocol, in accordance with~\cite{Sugimoto_00, Yan_08, Ii-Yung_13}, and (ii) for frames of length $10^{4}$ bits, the bit error rate after the second pass suggests block sizes for the third and subsequent passes larger than half of the frame length. As shown in Fig.~\ref{fig:bit-error-rate}, after the second pass the bit error rate is very low and its inverse is larger than $n/2$ (where $n$ is the frame length). In this case, the use of smaller blocks reveals many parities corresponding to blocks without errors. Note that, the number of parities disclosed for detecting and correcting errors during the $i$-th pass is approximated by $\lceil n/k_{i} \rceil + b_{i} \log_{2} k_{i}$, where $b_{i}$ is the number of blocks with parity mismatch for which a binary search is performed. Thus, assuming that only two errors remain in the frame and those are detected and corrected, a block size of $n/2$ is approximately optimal. This last conclusion also somehow agrees with the proposal in~\cite{Sugimoto_00}; however, for reasons not explained in that paper, the authors in~\cite{Sugimoto_00} use \BICONF for further passes instead of continuing taking advantage of the error backtracking feature of \Cascade to correct further errors in previous passes.

Unfortunately, no clear criterion for the second block size $k_{2}$ can be extracted. We might mistakenly infer that the optimal value for the second block size should be calculated similarly to the first block size. We can calculate this size using the expected error rate after the first pass as proposed in~\cite{Brassard_94}. Adopting the same notation, let $k_{1}=\lceil 1/Q \rceil$ and $E_{1}$ be the expected number of errors in a block after the completion of the first pass, we get:

\begin{equation}
E_{1} = \frac{1+(1-2Q)^{\lceil 1/Q \rceil}}{2}.
\end{equation}

Therefore, the block size for the second pass of \Cascade will be $k_{2}=k_{1}/E_{1}$ in order to optimize the number of errors that can be corrected during that pass. This size corresponds to approximately $k_{2} \approx 1.8 k_{1}$. However, simulations with these parameters quickly show that the efficiency in fact worsens. This is because the assumption above ignores the backtracking error correction carried out by \Cascade: while larger block sizes are less able to correct errors, the cascade effects more than compensate this and efficiently corrects errors using the block from the first pass. For this reason we chose to use the original protocol rule for selecting the second block size.

In summary, we propose here a first optimized version of \Cascade with the following parameters: $k_{1}=1/Q$, $k_{2}=2k_{1}$ and $k_{i}=\lceil n/2 \rceil$ for $i>2$; where the number of passes $i$ depends on the target frame error rate. As shown below, this initial approach is already closer to being optimal than any of the previous proposals.

\begin{figure}[ht]
\centering
\includegraphics[width=0.8\linewidth]{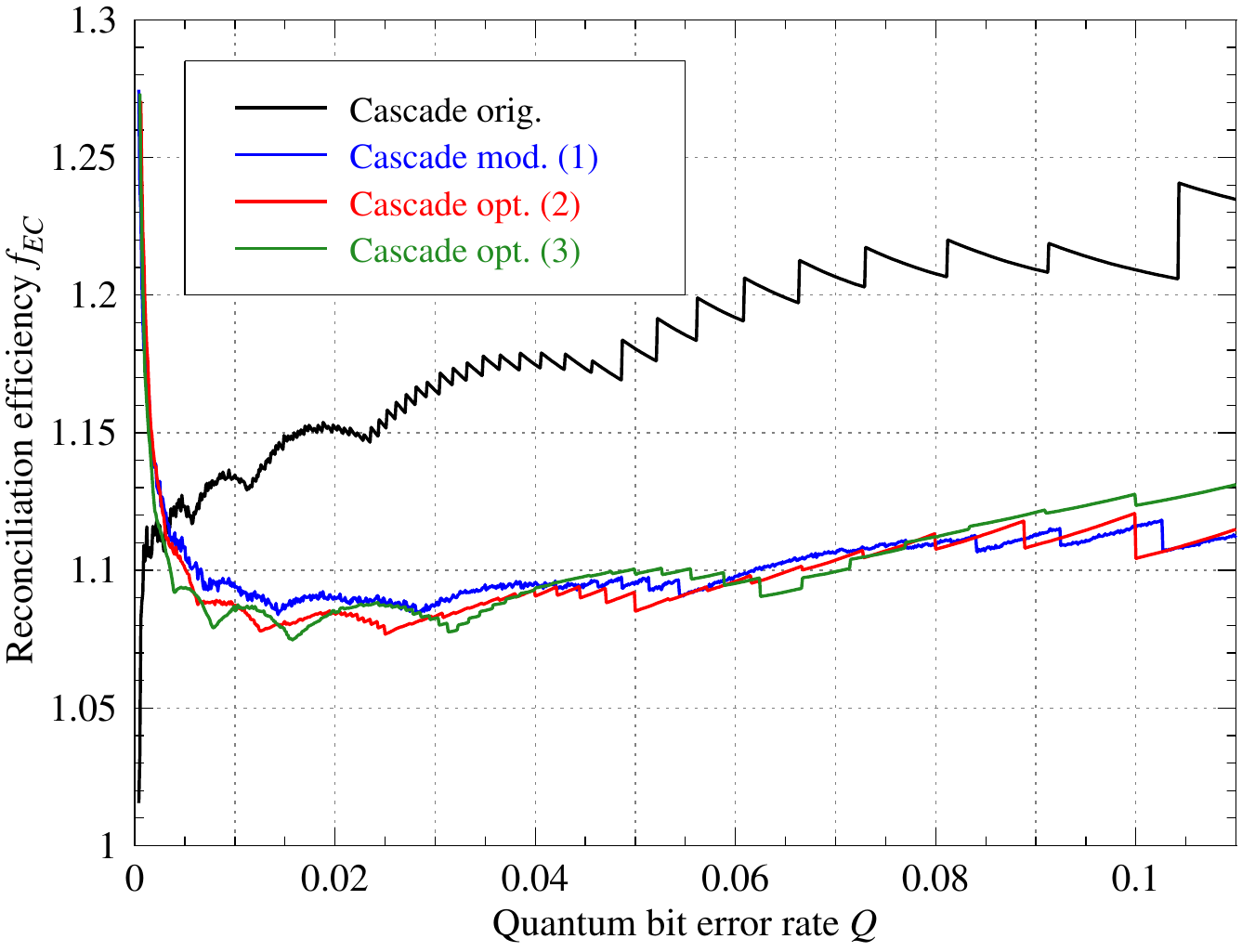}
\caption{Average reconciliation efficiency, $f_{EC}$, of the original \Cascade (black) and three modified versions: (1) the modified protocol proposed in~\cite{Sugimoto_00} (blue), (2) the version using the optimized parameters suggested in~\cite{Yan_08} (red), and (3) the version proposed here using $16$ passes (green). More details are given in the text.}
\label{fig:efficiency-opt}
\end{figure}

Fig.~\ref{fig:efficiency-opt} shows the average reconciliation efficiency as a function of QBER. Results were computed again for frames of length $n=10^{4}$ bits. The original \Cascade (black) is compared to three modified versions: (1) the modified protocol proposed in~\cite{Sugimoto_00} combining the first two passes of \Cascade with \texttt{BICONF}($10$) (blue), (2) the version using the optimized parameters suggested in~\cite{Yan_08}, i.e., $k_{1}=0.8/Q$, $k_{2}=5k_{1}$ and $k_{i}=n/2$ for $2<i\le 10$ (red), and (3) the version using the parameters proposed above and carrying out 16 passes (green). As shown, the efficiency is similar in the three proposed optimizations, despite using different block sizes for the first and second passes. It corresponds approximately to closing half of the gap between the efficiency of the original \Cascade and the theoretically optimal efficiency.

\begin{figure}[ht]
\centering
\includegraphics[width=0.8\linewidth]{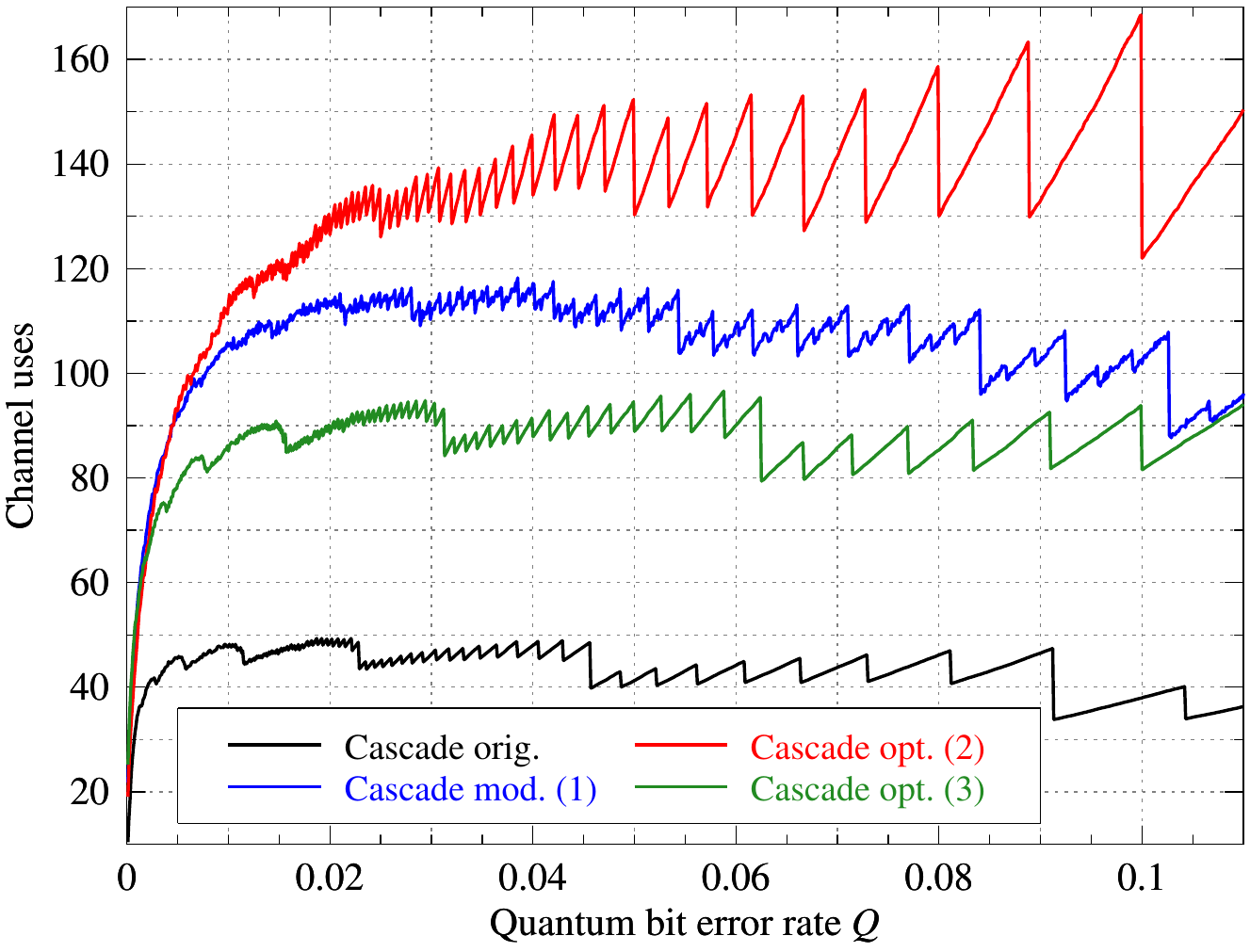}
\caption{Channel uses of the original \Cascade (black) and three modified versions: (1) the modified protocol proposed in~\cite{Sugimoto_00} (blue), (2) the version using the optimized parameters suggested in~\cite{Yan_08} (red), and (3) the version proposed here using $16$ passes (green). More details are given in the text.}
\label{fig:channel-uses-opt}
\end{figure}

Next, Fig.~\ref{fig:channel-uses-opt} shows the number of channel uses as a function of QBER for the four cases considered in Fig.~\ref{fig:efficiency-opt}. As shown, all the optimizations exceed the number of communication rounds of the original protocol, but the one proposed here shows the smallest number of channel uses of all three alternatives despite having used 16 passes.

\begin{figure}[ht]
\centering
\includegraphics[width=0.8\linewidth]{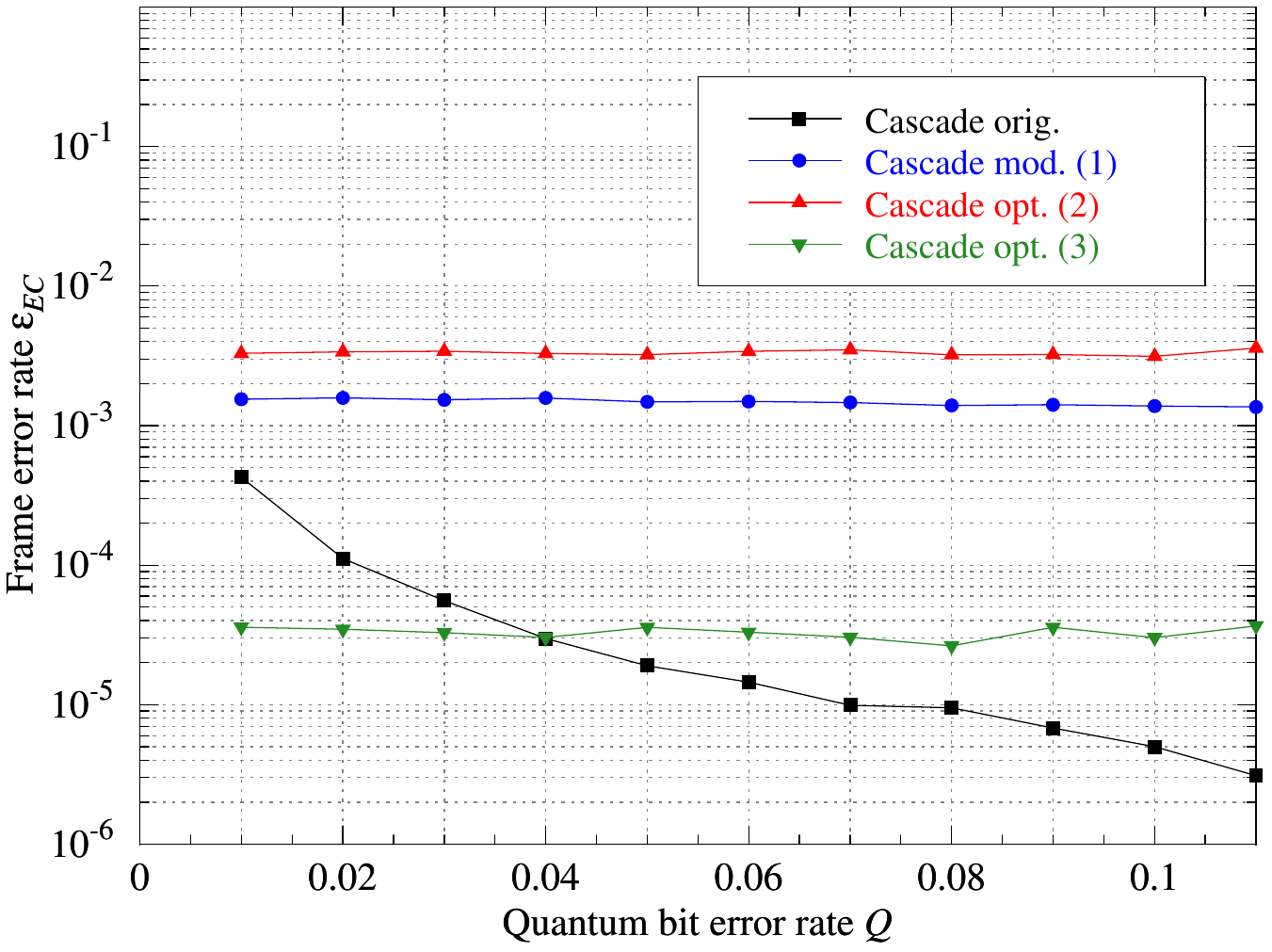}
\caption{Frame error rate (failure probability) of the original \Cascade (black) and three modified versions: (1) the modified protocol proposed in~\cite{Sugimoto_00} (blue), (2) the version using the optimized parameters suggested in~\cite{Yan_08} (red), and (3) the version proposed here using $16$ passes (green). More details are given in the text.}
\label{fig:frame-error-rate-opt}
\end{figure}

Finally, the frame error rate is shown in Fig.~\ref{fig:frame-error-rate-opt} for the four cases considered in the previous figures. These results show that, similarly to the other two \Cascade optimizations, the parameters proposed here also achieve a frame error rate independent of the QBER, which is however smaller by more than one order of magnitude compared to the frame error rate achieved by the previous two optimizations, and which is comparable to the average frame error rate of \Cascade. Note that, from the third pass onward, the frame error rate in the three studied optimizations decreases with the number of passes approximately as $2^{-s}$, where $s$ is the number of passes executed with a block size of half of the frame length. Curiously, this block length choice was suggested in~\cite{Sugimoto_00, Yan_08} although without further justification and somewhat in contradiction with the claims in these publications: the inferred frame error rate of zero gives no motivation for the number of passes suggested in both protocols. Later we discuss how the frame error rate influences the protocol and we try to justify the optimal number of passes that the algorithm must perform to achieve the best performance.

\subsection{Further optimized implementations of \Cascade}

Apart from the optimization of block sizes, we also analyze further optimization in the implementation of \Cascade. In this respect we utilize optimization approaches outlined in Section~\ref{sec:opt-parameters}.

\begin{figure}[ht]
\centering
\includegraphics[width=0.8\linewidth]{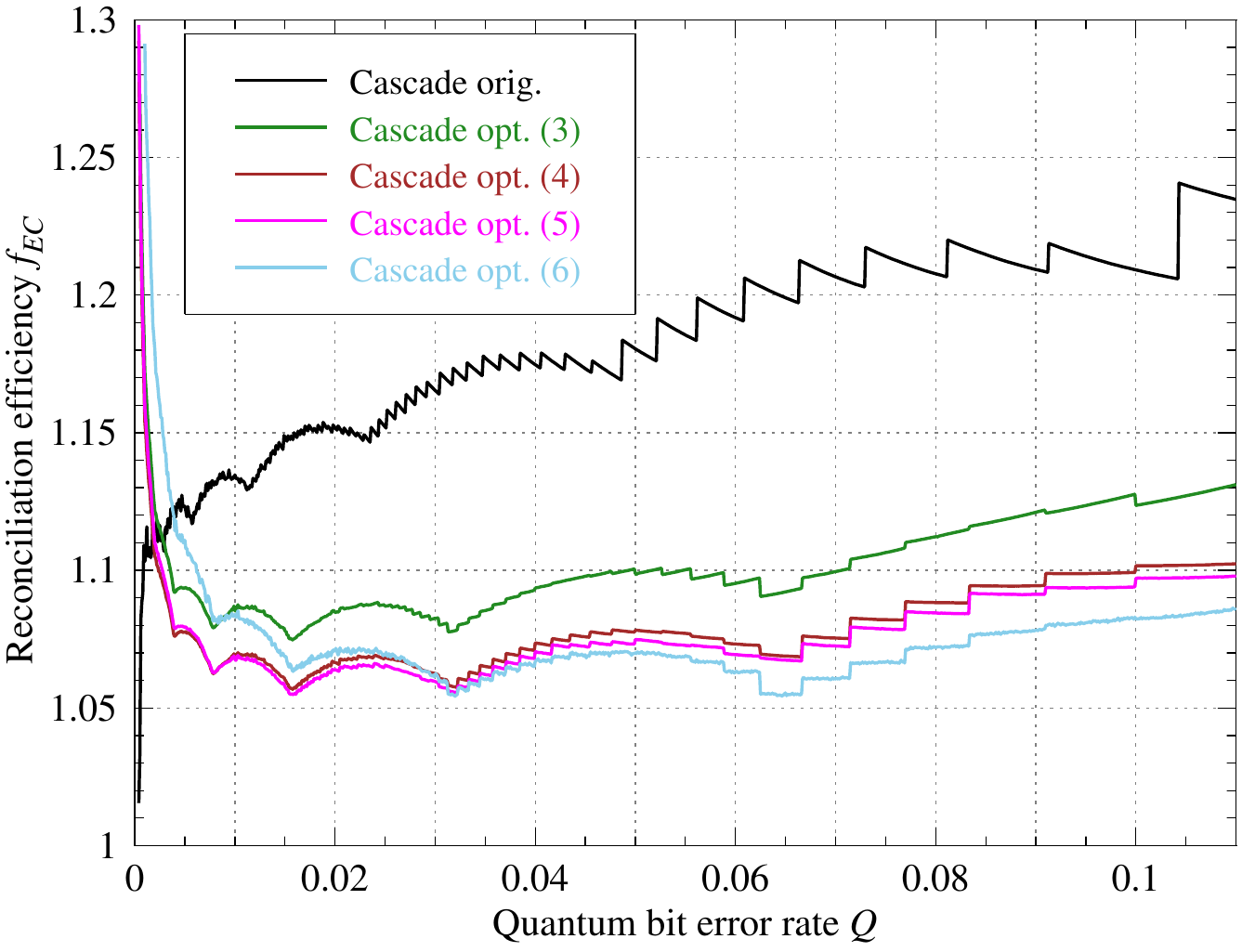}
\caption{Average reconciliation efficiency, $f_{EC}$, of the original \Cascade (black) and for optimized versions: (3) the version using $16$ passes, proposed above and presented in the previous figures (green), (4) same as (3) but leveraging in addition the idea of block reuse as suggested by~\cite{Yan_08} (brown), (5) same as (4) but replacing the random shuffling between passes (magenta), and (6) same as (3) but discarding singleton blocks after each pass (sky blue). More details are given in the text.}
\label{fig:efficiency-opt-2}
\end{figure}

Fig.~\ref{fig:efficiency-opt-2} presents the efficiency as a function of QBER for several implementations in addition to an implementation of the original \Cascade (black curve). The first optimization approach, that we have put forward and analyzed in detail on the basis of simulation results above, corresponds to the green curve, labeled with (3). First we compare it to the improvement in efficiency arising from block reuse as proposed in~\cite{Yan_08}. This approach is the basis for a further implementation of \Cascade labeled with (4) (brown curve), which uses a record of all processed blocks per pass and the optimized parameters suggested above. From the figure it is clear that this implementation leads to a significant increase of the efficiency, which comes at the cost of only higher memory usage, since pointers to all subblocks have to be kept, and a more complicated implementation\footnote{The number of communication rounds and frame error rate coincide for these two optimizations, labeled with (3) and (4).}. Note that, in~\cite{Yan_08} it is suggested, first to sort the list of subblocks by size, and then to correct the shortest one. However, we have implemented a different version in which all the subblocks in a pass are processed in parallel regardless of their size. Thus, although the efficiency might worsen a bit, we are not penalizing the number of communication rounds.

Fig.~\ref{fig:efficiency-opt-2} also shows the efficiency of two further optimized implementations, labeled with (5) and (6), respectively. These make use of the approaches put forward in~\cite{Nguyen_02, VanAssche_05, VanAssche_06, Yan_08}. The curve labeled with (5) and colored in magenta is the result of replacing the random shuffling between passes in the implementation labeled with (4) by an improved one. Note that the efficiency depicted in the figure is again the result of a slightly different interpretation of an improved shuffling in comparison to the original proposals. Thus, instead of using a deterministic shuffling, as proposed in~\cite{Nguyen_02}, we continue using a random one to avoid that two bits of a subblock might coincide in the same block of a subsequent pass. As shown, the modified shuffling marginally improves the efficiency of the implementation labeled with (4). The other curve labeled with (6) is obtained by discarding the singleton blocks in successive passes as proposed in~\cite{Chen_01}. As shown, although the efficiency improves in the high QBER region, for low error rates it worsens. However, for a fair comparison, the second block size has to be adjusted for this optimization, given that the per block error probability after the first pass changes.

\subsection{Near optimal \Cascade parameters}

Up to now, efficiency and frame error rate have been considered separately. We have also seen that this can be dangerous, since it does not make sense to have a very high efficiency when actually many frames are discarded because of a high frame error rate. Hence, a better measure of the quality of the protocol would be a modified efficiency that takes into account the frame error rate. Further, we will justify the number of passes carried out in \Cascade based on this efficiency.

We define then the ratio of information leakage of an error reconciliation protocol as follows:

\begin{equation}
\mathrm{leak}_{EC} = (1 - \varepsilon_{EC}) (1 - R) + \varepsilon_{EC}
\label{eq:leakage}
\end{equation}

\noindent where $\varepsilon_{EC}$ is the frame error rate in the reconciliation procedure, and $R$ is the ratio of information transmitted, as defined in Eq.~(\ref{eq:efficiency-bsc}). The factors $1-\varepsilon_{EC}$ and $1-R$ correspond to the probability of successful reconciliation of two frames and the ratio of information disclosed for reconciling errors in the frames, respectively. Note that in this definition of leakage we implicitly assume that the entire frame is disclosed when the reconciliation procedure fails, or equivalently that in case of error frames are discarded. In this way reconciliation is always guaranteed in a simulation context. Thus, although this definition penalizes the efficiency, it has the advantage of not having to consider frame error rate explicitly.

\begin{figure}[ht]
\centering
\includegraphics[width=0.8\linewidth]{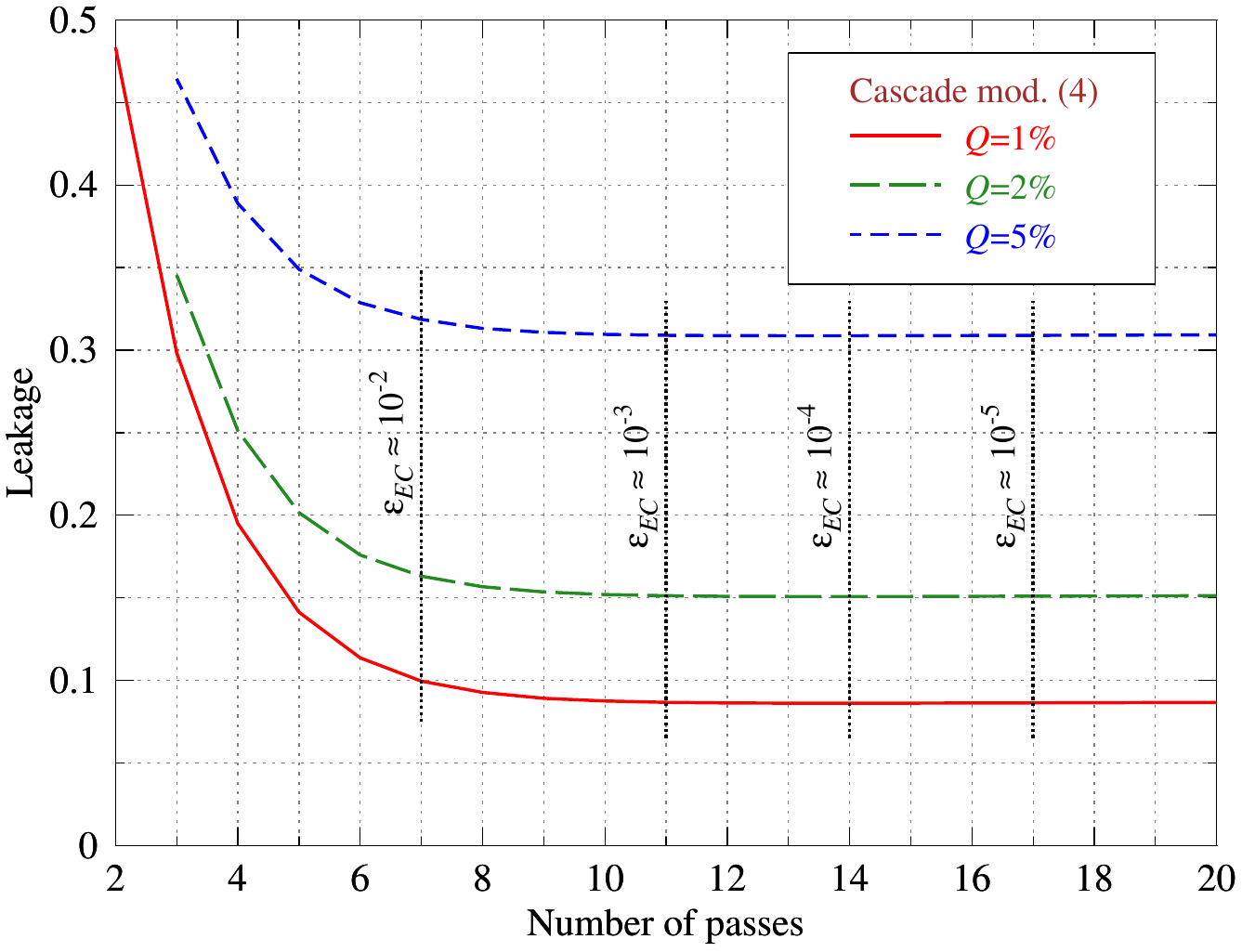}
\caption{Information leakage as a function of the number of passes in the proposed modification of \Cascade utilizing subblock reuse, labeled with (4) in Fig.~\ref{fig:efficiency-opt-2}.}
\label{fig:leakage-passes}
\end{figure}

Fig.~\ref{fig:leakage-passes} shows the information leakage, as described in Eq.~(\ref{eq:leakage}), as a function of the number of passes for the modified version of \Cascade proposed here, utilizing subblock reuse, i.e., the one labeled with (4) above. Three different QBER values, $Q=1\%$, $2\%$ and $5\%$, are considered. Approximate values for the frame error rate after completing several passes are also marked in the figure, from left to right, $\varepsilon_{EC} \approx 1.6 \times 10^{-2}$ after completing 7 passes, $10^{-3}$ after 11 passes, $1.2 \times 10^{-4}$ after 14 passes and $1.6 \times 10^{-5}$ after 17. Although 16 passes have been carried out to fairly compare the proposed modification of \Cascade with the original one, it is clear from this figure that between 10 and 12 passes are enough to achieve the near optimal leakage of the protocol. Note that the frame error rate after these passes strikingly corresponds to the one of the optimized protocol proposed in~\cite{Sugimoto_00, Yan_08}. However, for the three QBER values simulated, the optimum is obtained in all the cases for 14 passes.

Henceforth, we use this leakage definition to provide a description of the efficiency that takes into account the frame error rate as follows. As in Section~\ref{sec:pre} we use $\epsilon$ for the QBER and $h(\epsilon)$ for the binary Shannon entropy. The reconciliation efficiency is given by:

\begin{equation}
\eta_{EC} = \frac{\mathrm{leak}_{EC}}{h(\epsilon)}.
\label{eq:efficiency-eta}
\end{equation}

We use Eq.~(\ref{eq:efficiency-eta}) to optimize the first and second block sizes for the range of QBER considered here. In order to find the optimal block sizes that minimize the reconciliation efficiency we use a Compass search algorithm (a simple case of a generating set search method) \cite{Conn_09}. This is a two dimensional direct search algorithm that allows minimizing a function without calculating derivatives, hence very robust and reliable. This works as follows. Firstly, it chooses initial values for the variables to optimize and a delta value for the step size, e.g., our choices have been $k_{1}=1/Q$, $k_{2}=2k_{1}$ and $\delta=k_{1}$. Then, it minimizes the function to get the efficiency $\eta_{\min}$ for the two initial block sizes, $k_{1}$ and $k_{2}$. In each iteration the Compass search algorithm computes the function to minimize for four possible directions: North, South, East, West; i.e., it computes the efficiency for the following four alternatives: $(k_{1}+\delta,k_{2})$, $(k_{1}-\delta,k_{2})$, $(k_{1},k_{2}+\delta)$ and $(k_{1},k_{2}-\delta)$. If the best of the computed efficiencies improves $\eta_{\min}$, the algorithm updates the block sizes and the minimum efficiency with the best values. If none of these efficiencies improve the current one, the delta value is decreased by $20\%$, i.e., $\delta=4\delta/5$, and a new iteration begins.

\begin{table}[ht]
\centering
\caption{Optimized values for the first and second block sizes using a Compass search algorithm.}
\label{tab:opt-block-sizes}
\begin{tabular}{|c|c|ccccccccccc|}
\hline
$n$ & $Q$ & $1\%$ & $2\%$ & $3\%$ & $4\%$ & $5\%$ & $6\%$ & $7\%$ & $8\%$ & $9\%$ & $10\%$ & $11\%$ \\
\hline
\multirow{2}{*}{$10^{4}$} & $k_{1}$ & 125 & 64 & 32 & 32 & 32 & 16 & 16 & 16 & 16 & 16 & 16 \\
 & $k_{2}$ & 400 & 250 & 172 & 128 & 64 & 67 & 64 & 63 & 64 & 63 & 65 \\
\hline
\multirow{2}{*}{$2^{14}$} & $k_{1}$ & 128 & 64 & 32 & 32 & 32 & 16 & 16 & 16 & 16 & 16 & 16 \\
 & $k_{2}$ & 520 & 256 & 128 & 128 & 128 & 64 & 64 & 64 & 64 & 64 & 64 \\
\hline
\end{tabular}
\end{table}

Table~\ref{tab:opt-block-sizes} shows the optimized values for the first and second block sizes obtained using the Compass search algorithm described above. Results are given for different QBER values. These results have been initially computed for frames of length $10^{4}$ bits, using simulations comprising $10^{4}$ frames for each point, a high enough number to get a reasonably accurate idea of the optimal block sizes. For this optimization the improved version of \Cascade that includes only an implementation of the subblock reuse, labeled with (4), has been employed using the same number of passes (i.e., 14 passes). As shown in the table, the optimal efficiency is obtained most of the time for $k_{1}$ and $k_{2}$ values that are powers of two or nearby values. Note that as the block sizes move away from numbers that are a power of two, the dichotomic search tends to produce increasingly subblocks of size 3, that work inefficiently. Consequently, results were later computed with a higher accuracy ($10^{5}$ frames per point) to look for the block sizes that optimize the efficiency of a power of two frame length $n=2^{14}$. The results for this frame length are also presented in Table~\ref{tab:opt-block-sizes}. They show, even more convincingly, the importance of using power of two block sizes. In fact, the use of power of two block sizes is even more important than any other protocol optimizations to improve the average reconciliation efficiency of \Cascade.

\begin{table}[ht]
\centering
\caption{Optimized values for the first, second and third block sizes using a Compass search algorithm, average reconciliation efficiencies, and channel uses. Note how the frame error rate is kept almost constant and close to $10^{-4}$.}
\label{tab:opt-block-sizes-2}
\begin{tabular}{|c|r|r|r|l|l|l|l|c|}
\hline
 & & & & & & & & Chan. \\
$Q$ & \multicolumn{1}{|c|}{$k_{1}$} & \multicolumn{1}{|c|}{$k_{2}$} & \multicolumn{1}{|c|}{$k_{3}$} & \multicolumn{1}{|c|}{$\eta_{EC}$} & \multicolumn{1}{|c|}{$f_{EC}$} & \multicolumn{1}{|c|}{$\varepsilon_{EC}$} & \multicolumn{1}{|c|}{$\beta$} & uses \\
\hline
$0.5\%$ & 256 & 1024 & 4096 & $1.05182$ & $1.04989$ & $9.2 \times 10^{-5}$ & $0.9976$ & $168.6$ \\
$1\%$   & 128 &  512 & 4096 & $1.0431$  & $1.04219$ & $8.0 \times 10^{-5}$ & $0.9963$ & $208.8$ \\
$2\%$   &  64 &  512 & 4096 & $1.04062$ & $1.04006$ & $9.3 \times 10^{-5}$ & $0.9934$ & $407.6$ \\
$3\%$   &  32 &  512 & 4096 & $1.03945$ & $1.03902$ & $1.1 \times 10^{-4}$ & $0.9906$ & $496.9$ \\
$4\%$   &  32 &  256 & 4096 & $1.04342$ & $1.04313$ & $9.4 \times 10^{-5}$ & $0.9862$ & $500.2$ \\
$5\%$   &  16 &  256 & 4096 & $1.04335$ & $1.04313$ & $8.9 \times 10^{-5}$ & $0.9827$ & $432.6$ \\
$6\%$   &  16 &  256 & 4096 & $1.04601$ & $1.0458$  & $1.1 \times 10^{-4}$ & $0.9777$ & $606.6$ \\
$7\%$   &  16 &  256 & 4096 & $1.05065$ & $1.0505$  & $8.7 \times 10^{-5}$ & $0.9709$ & $796.9$ \\
$8\%$   &   8 &  256 & 4096 & $1.05479$ & $1.05465$ & $9.7 \times 10^{-5}$ & $0.9632$ & $550.3$ \\
$9\%$   &   8 &  256 & 4096 & $1.05499$ & $1.05486$ & $1.0 \times 10^{-4}$ & $0.9575$ & $690.4$ \\
$10\%$  &   8 &  256 & 4096 & $1.05747$ & $1.05736$ & $1.0 \times 10^{-4}$ & $0.9493$ & $840.3$ \\
$11\%$  &   8 &  256 & 4096 & $1.06139$ & $1.0613$  & $1.0 \times 10^{-4}$ & $0.9387$ & $998.4$ \\
\hline
\end{tabular}
\end{table}

A search for the optimal block sizes, also considering the third block size $k_{3}$, is then carried out. To reduce the complexity of the search, and since we already know that blocks that are not a power of two are not going to be optimal, the search only considers power of two subblocks, thus reducing significantly the amount of heavy calculations needed. The frame length also corresponds to a power of two as in Table~\ref{tab:opt-block-sizes}, $n=2^{14}=16384$. Table~\ref{tab:opt-block-sizes-2} shows the optimal block sizes achieved for different QBER values. For an easier comparison with previous results, in the table the average reconciliation efficiency, as described in Eqs.~(\ref{eq:efficiency-bsc}), (\ref{eq:efficiency-beta}) and~(\ref{eq:efficiency-eta}), and the number of channel uses are also included. Note that a significant price to pay for the improvement of the efficiency is in the number of communication rounds.

As a result of this analysis we propose to use the following near optimal parameters in \Cascade: $k_{1} = 2^{\lceil \alpha \rceil}$, $k_{2} = 2^{\lceil (\alpha + 12)/2 \rceil}$, $k_{3} = 4096$, and $k_{i}=\lceil n/2 \rceil$, where $\alpha = \log_{2} 1/Q - \frac{1}{2}$ and a frame length of $n=2^{14}$; and to optimize the protocol implementation by considering the suggested subblock reuse.

\begin{figure}[ht]
\centering
\includegraphics[width=0.8\linewidth]{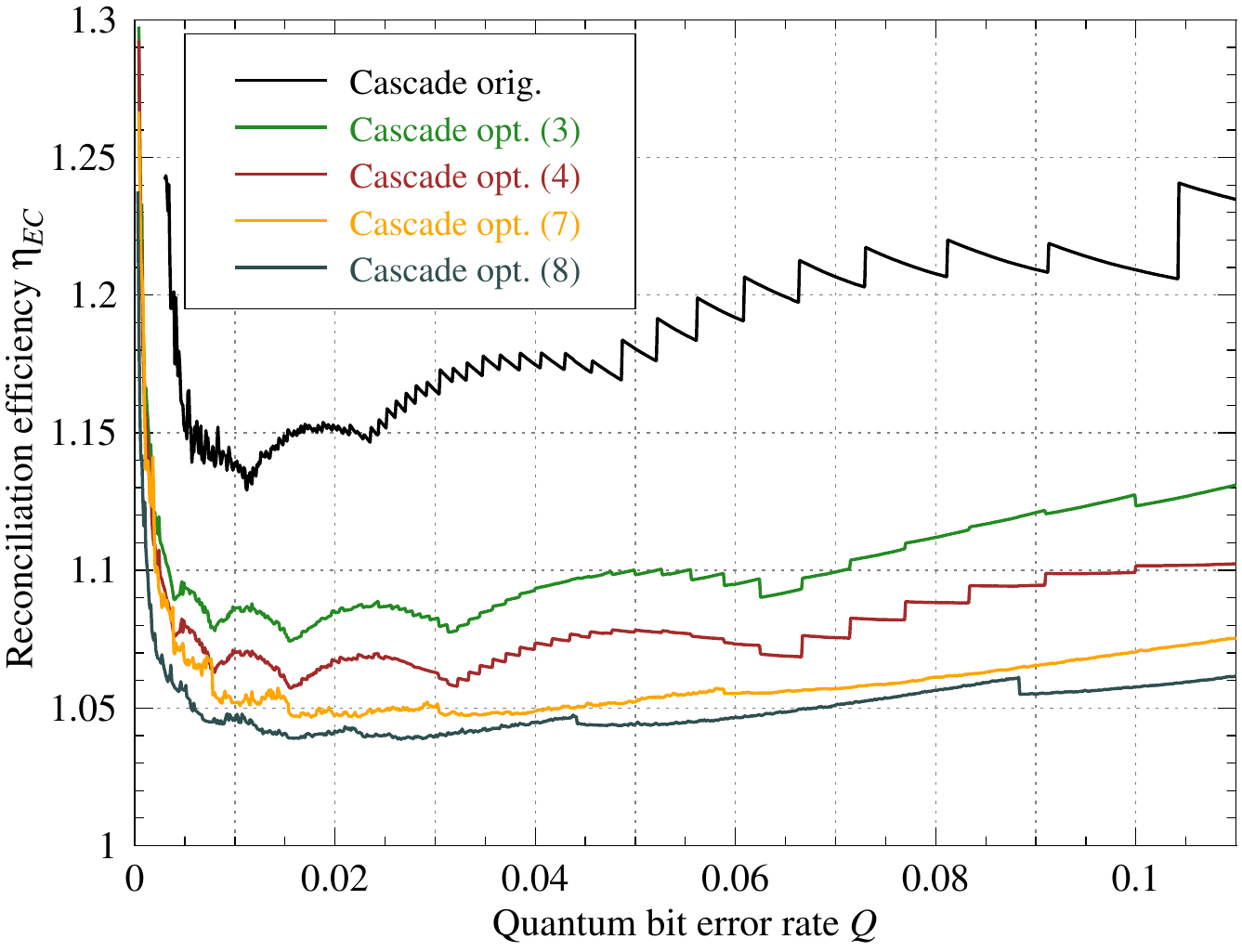}
\caption{Average reconciliation efficiency, $\eta_{EC}$, of the original \Cascade (black) and for optimized versions: (3) the first version proposed above and presented in the previous figures using $16$ passes (green), (4) same as (3) but leveraging in addition the idea of block reuse (brown), (7) same as (4) but optimizing the first and second block sizes and using 14 passes (orange), and (8) same as (7) but also optimizing the third block size and using a power of two value for the frame length $n=2^{14}$ (dark gray). More details are given in the text.}
\label{fig:efficiency-opt-3}
\end{figure}

Fig.~\ref{fig:efficiency-opt-3} shows the average reconciliation efficiency for the two optimized implementations labeled with (3) and (4) in Fig.~\ref{fig:efficiency-opt-2}. Efficiency is now calculated using Eq.~(\ref{eq:efficiency-eta}) to take into account the frame error rate. As shown, the efficiency does not decrease to one in the low QBER region, but it increases due to the contribution of the frame error rate. Curves are then fairly comparable among different optimizations, in particular, for the low error rate region. Now, as expected, the efficiency goes to infinity for all the curves when the error rate tends to zero: even disclosing only one parity, if the error is close to zero brings about very high increase in efficiency. In the figure two additional optimized versions of \Cascade labeled with (7) and (8) are also included. The curve labeled with (7) is the result of optimizing the first and second block sizes with the suggested power of two values of Table~\ref{tab:opt-block-sizes}, i.e., $k_{1} = 2^{\lceil \log_{2} 1/Q \rceil}$, $k_{2}=4k_{1}$, $k_{i}=\lceil n/2 \rceil$ for $i>2$ and $n=10^{4}$, with 14 passes. The curve labeled with (8) is the result of optimizing the first, second and third block sites as suggested in Table~\ref{tab:opt-block-sizes-2} and using a frame length of $n=2^{14}$. As far as we know these are the best efficiency values obtained with \Cascade or any of its modifications. Furthermore, these values are not unrealistic, since they take into account the frame error rate. Note that this implies a rather high number of communications. This is an issue that is likely to be of importance for high speed QKD systems working at low QBER regimes, where the classical post-processing can be the bottleneck for the final secret key throughput. However, for long distance, high losses, QKD transmissions, where every extra secret bit counts, this is likely to be a minor issue. Obviously, the user can also choose to implement some other of the proposed modifications to get good efficiency and low frame error rate but with a reduced communication cost. For example, optimizations (3) and (4) strike a good balance between efficiency, frame error rate and communication cost. Optimization (4) is slightly more efficient than (3) for the same frame error rate and communication cost, but the implementation complexity and required hardware resources are higher.

\section{Conclusions}
\label{sec:conclusions}

We provide a comprehensive comparison of the \Cascade information reconciliation protocol and some of its modified versions that have been proposed in literature. Results of exhaustive simulation studies have been used to compare the efficiency, communication rounds and robustness (failure probability or frame error rate) for all discussed versions. It is shown that simple claims like efficiency improvement alone do not justify the adoption of a particular modification. A more global view is needed and, in particular, the frame error rate has to be taken into account. Based on the analysis of our results, we also propose an optimized version of \Cascade that utilizes previous ideas, and leads to a near optimal implementation of the protocol. Our optimization is based on reconciling frames with lengths of $10^{4}$ bits, and although it is also partly valid for larger frames, to achieve optimal performance the block sizes should be newly optimized. Preliminary calculations indicate that larger frame lengths will further improve the average reconciliation efficiency, albeit marginally. It is shown that this optimization, when used with frames that are a power of two, achieves a record reconciliation efficiency while keeping the frame error rate low.

\section*{Acknowledgment}

The first author wishes to thank Thomas B. Pedersen for helpful discussions about the \Cascade implementation and some results. The authors also gratefully acknowledge the computer resources, technical expertise and assistance provided by the \emph{Centro de Supercomputaci\'{o}n y Visualizaci\'{o}n de Madrid}\footnote{http://www.cesvima.upm.es} (CeSViMa).

This work has been partially supported by the project Hybrid Quantum Networks, TEC2012-35673, funded by \textit{Ministerio de Econom\'{i}a y Competitividad}, Spain and by the Vienna Science and Technology Fund (WWTF) through project ICT10-067 (HiPANQ).

\bibliographystyle{unsrtnat}
\bibliography{cascade}

\begin{thebibliography}{32}
\bibitem[Mom()]{Momtchil_14}
This is a crucial issue for the application of \Cascade in QKD. Information
  leakage, e.g., see Eq.~(\ref{eq:leakage}), is a quantity to be subtracted
  from the raw key in the privacy amplification step of the post-processing
  protocol. Information leakage in turn depends on the ratio of the transmitted
  information and hence on the information efficiency $f_{EC}$ (cf.
  Eq.~(\ref{eq:efficiency-bsc}); the theoretically optimal value of the latter
  being equal to 1). The main question then is how to estimate the information
  leakage of \Cascade. Our approach here is that if a parity of a subblock of
  one of the parties (Alice) is sent to the other (Bob) and he responds with
  the parity of the same subblock, then the overall leakage is 1 bit. This is
  by no means self-evident as in fact two bits are leaked and in principle the
  eavesdropper could make use of both of them. If so, the leakage would double
  and the usage of \Cascade and any two-way protocol in general would be highly
  penalized during privacy amplification. Fortunately this is not the case, at
  least for BB84. In~\cite{Lo_03} it is shown that the information leakage in
  the discussed case is indeed 1 bit for arbitrary eavesdropper attacks. The
  author puts forward the approach of encrypting the reconciliation
  communication exchange by one time pad, a procedure that consumes key and is
  thus, from an overall key generation balance point of view, equivalent to
  reducing an equal amount of key during privacy amplification. One of the
  results in~\cite{Lo_03} is that for BB84 one can encrypt BOTH parities
  discussed above by the same bit. This implies that in this case the
  information leakage due to Bob's answer does not need to be taken into
  account during privacy amplification and it is the leakage of information,
  due to Alice's communication alone that is relevant for \Cascade performance
  for BB84.
\end{thebibliography}

\end{document}